\documentclass[aps,prb,superscriptaddress,notitlepage,showpacs]{revtex4-1}

\usepackage{natbib}
\usepackage{hyperref}
\hypersetup{
    colorlinks=true,
}

\usepackage{mathtools}
\usepackage{amsfonts}
\usepackage{amssymb}

\usepackage{graphicx}
\usepackage[all]{xy}

\usepackage{xspace}
\usepackage{amsthm}

\newcommand{\Hpp}{H_{pp}}
\newcommand{\Hpq}{H_{pq}}
\newcommand{\Hpqqr}{H_{pqqr}}
\newcommand{\Hprrq}{H_{prrq}}
\newcommand{\Hpqqp}{H_{pqqp}}
\newcommand{\Hpqrs}{H_{pqrs}}

\newcommand{\lqd}{LIQ$Ui|\rangle$\xspace}
\newcommand{\be}{\begin{equation}}
\newcommand{\ee}{\end{equation}}

\begin{document}
\title{Improving Quantum Algorithms for Quantum Chemistry}
\author{M. B. Hastings}
\affiliation{Station Q, Microsoft Research, Santa Barbara, CA 93106-6105, USA}
\affiliation{Quantum Architectures and Computation Group, Microsoft Research, Redmond, WA 98052, USA}

\author{Dave Wecker}
\affiliation{Quantum Architectures and Computation Group, Microsoft Research, Redmond, WA 98052, USA}

\author{Bela Bauer}
\affiliation{Station Q, Microsoft Research, Santa Barbara, CA 93106-6105, USA}

\author{Matthias Troyer}
\affiliation{Theoretische Physik, ETH Zurich, 8093 Zurich, Switzerland}

\begin{abstract}
We present several improvements to the standard Trotter-Suzuki based algorithms used in the simulation of quantum
chemistry on a quantum computer.
First, we modify how Jordan-Wigner transformations are implemented to reduce their cost from linear or logarithmic in
the number of orbitals to a constant. Our modification does not require additional ancilla qubits.
Then, we demonstrate how many operations can be parallelized, leading to
a further linear decrease in the parallel depth of the circuit,
at the cost of a small constant factor increase in number of qubits required. Thirdly, we modify the term order in the Trotter-Suzuki decomposition, significantly reducing
the error at given Trotter-Suzuki timestep. A final improvement modifies the Hamiltonian to reduce errors introduced by the non-zero Trotter-Suzuki timestep.
All of these techniques are validated using numerical simulation and detailed gate counts are given for realistic molecules.
\end{abstract}
\maketitle

One of the most natural applications of a quantum computer is to simulate quantum mechanics, as suggested
by Feynman~\cite{Feynman1982}. There has been much work on constructing quantum algorithms to simulate
various problems in quantum mechanics, ranging from problems in condensed matter physics such as the
Hubbard model to quantum field theory~\cite{qft} and quantum
chemistry~\cite{lidar1999,ortiz2001,AspuruGuzik2005,kassal2008,wang2008,kassal2009,Lanyon2010,whitfield2011,kassal2011,yung2012introduction,whitfield2013spin,yung2013,lamata2013,toloui2013}.

A recent large-scale study of quantum chemistry on a quantum computer~\cite{qchem} gave accurate gate counts for some of the standard circuits in the literature when applied to problems in quantum chemistry. Unfortunately, this study found that even modest molecules require enormously long simulation time. The reason for this is simple: in quantum chemistry, we consider a basis with $N$ spin orbitals. The
Hamiltonian considered takes the form
\begin{equation} \label{eqn:H}
H = \sum_{pq} t_{pq} c_p^\dagger c_q + \frac{1}{2}\sum_{pqrs} V_{pqrs} c_p^\dagger c_q^\dagger c_r c_s
\end{equation}
for a system of interacting electrons. Many of the two-body interaction terms $V_{pqrs}$ are non-zero and so there are roughly $N^4$ separate, non-commuting terms in the Hamiltonian.  The time requirements become large even for roughly 100 spin orbitals; since roughly 50-70 spin orbitals can already be simulated on a classical computer using exact or approximate techniques~\cite{nakano2011,lauchli2011ground,capponi2013numerical,gan2005calibrating,Kurashige2013}, clearly improvements are needed in the quantum algorithms to make them potentially useful.

Typical quantum algorithms to simulate this system need to implement unitary time evolution under the Hamiltonian~(\ref{eqn:H}). Efficient ways to implement this evolution have thus been the object of intense research efforts~\cite{efficient2007,childs2011simulating,anmer2011,childs2012hamiltonian,childs2012product,berry2012black,anmer2012,berry2013exponential2}.
Other approaches that do not rely on time evolution have been proposed~\cite{babbush2013,peruzzo2013}. For the time evolution, a Trotter-Suzuki approach~\cite{trotter1959,suzuki1976} is most common, where the evolution $\exp(i H \delta_t)$ for a small time step $\delta_t$ (controlled by an additional ancilla qubit used to perform the phase estimation) is implemented through a sequence of $O(N^4)$ unitary transformations $\exp(i A \delta_t)$  where $A$ is some term in Eq.~(\ref{eqn:H}). Standard implementations~\cite{whitfield2011} have an additional factor of $N$ overhead in gate count to implement the Jordan-Wigner transformation, which encodes the anticommutation relations of the fermionic degrees of freedom, giving a time complexity $O(N^5)$.
Further, as $N$ increases, the Trotter time step must become smaller to obtain a fixed, given accuracy, further worsening the performance of the algorithm. All these effects combine to give poor scaling with $N$.

In this paper, we significantly alleviate these problems. One technique is a modified circuit which reduces the gate count overhead for Jordan-Wigner strings to a {\it constant} without requiring additional qubits. This improves both on the linear overhead of Ref.~\onlinecite{whitfield2011} and on the Bravyi-Kitaev scheme which has only a logarithmic overhead~\cite{BK}. Further, we show that with this modified circuit, many of the operations can be parallelized, leaving the total gate count unchanged but reducing the parallel circuit depth. All these improvements significantly reduce the gate count and parallel circuit depth both asymptotically and for small molecules.
Further, we modify the Trotter-Suzuki decomposition in two ways. First, we modify the term order to take into account special properties of a Hartree-Fock basis in quantum chemistry.
Second, we modify the Hamiltonian that we study in a way that corrects for having a nonzero Trotter step. These improvements allow us to obtain accurate results at much larger Trotter step than one might obtain otherwise.

These improvements then take two forms. One type of improvement involves modifying the circuits to perform the same simulation in a more effective way. The other type involves modifying the simulation done to obtain more accurate answers at larger Trotter step. Below, we detail all these improvements.
We explicitly check gate counts for various real molecules, using
\lqd, a quantum simulator developed at Microsoft Research~\cite{wecker2014}, so that we could obtain precise numbers for the improvements over the original circuits, rather than just asymptotic estimates.

To start, let us fix some terminology.  We say that a term in Eq.~(\ref{eqn:H}) is an $\Hpq$ term if it is of the form $t_{pq} c^\dagger_p c_q$ for $p\neq q$, while it is an $\Hpp$ term if $p=q$.  A term is an $\Hpqrs$ term if it is of the form $\frac{1}{2} V_{pqrs} c_p^\dagger c_q^\dagger c_r c_s$ for $p,q,r,s$ all distinct.  An $\Hpqqp$ term is of the form
$\frac{1}{2} V_{pqqp} c_p^\dagger c_q^\dagger c_q c_p$ while an $\Hpqqr$ term is of the form
$\frac{1}{2} V_{pqqr} c_p^\dagger c_q^\dagger c_q c_r$, again for $p,q,r$ all distinct.
Note that the $\Hpp$ and $\Hpqqp$ terms are diagonal.
Since $c_p c_q=-c_q c_p$ and similarly $c^\dagger_p c^\dagger_q = -c^\dagger_q c^\dagger_p$, we will rewrite any term
$\frac{1}{2} V_{pqpq} c_p^\dagger c_q^\dagger c_p c_q$  as an $\Hpqqp$ term and we will rewrite any term
$\frac{1}{2} V_{pqrq} c_p^\dagger c_q^\dagger c_r c_q$ as an $\Hpqqr$ term, while for an
an $\Hpqrs$ term we will assume that that $p<q$ and $r<s$.

For the molecules we studied, the terms with the largest magnitude are the $\Hpp$ terms. The $\Hpqqp$ terms are typically next in magnitude, then the $\Hpq$ terms, next the $\Hpqqr$ terms, and finally the $\Hpqrs$ terms. This ordering is not strict. Further, there is an important sum rule relating the $\Hpqqr$ and $\Hpq$ terms. After a basis transformation of the single-particle states to find a Hartree-Fock ground state, one has
\be
t_{pq}+\sum_r V_{prrq} n_r=0,
\ee
where $n_r=0,1$ is the occupation number in the Hartree-Fock state.

\section{Cancelling Jordan-Wigner Strings}
We work in a second quantized basis, using one qubit per spin orbital.
In the case of an $\Hpq$ term with $p<q$, the desired controlled unitary is 
\begin{equation}
\exp(i (\sigma^X_p \sigma^X_q + \sigma^Y_p \sigma^Y_q) (\sigma^Z_{p+1} \sigma^Z_{p+2} ... \sigma^Z_{q-1}) \theta),
\end{equation}
where the product of $Z$s implements the Jordan-Wigner string and where $\theta$ depends upon the coefficient $t_{pq}$ and on $\delta_t$. In the case of an $\Hpqrs$ term with $p<q<r<s$, there are several possible controlled unitaries, of the form
\begin{equation}
\label{pqrsunitary}
\exp(i (\sigma^X_p \sigma^X_q \sigma^X_r \sigma^X_s) (\sigma^Z_{p+1} ... \sigma^Z_{q-1}) (\sigma^Z_{r+1} ... \sigma^Z_{s-1}) \theta)
\end{equation} with Jordan-Wigner strings from $p+1$ to $q-1$ and from $r+1$ to $s-1$. In fact, for each $p,q,r,s$, it may be necessary to implement several of these terms, with $\sigma^X_p \sigma^X_q \sigma^X_r \sigma^X_s$ replaced by other Pauli matrices, such as $\sigma^X_p \sigma^X_q \sigma^Y_r \sigma^Y_s$, etc..., in every case having an even number of $\sigma^X$ operators and an even number of $\sigma^Y$ operators.  We refer to these different choices of $\sigma^X,\sigma^Y$ as different {\it subterms}.

\begin{figure}[t]
 \includegraphics[width=4in]{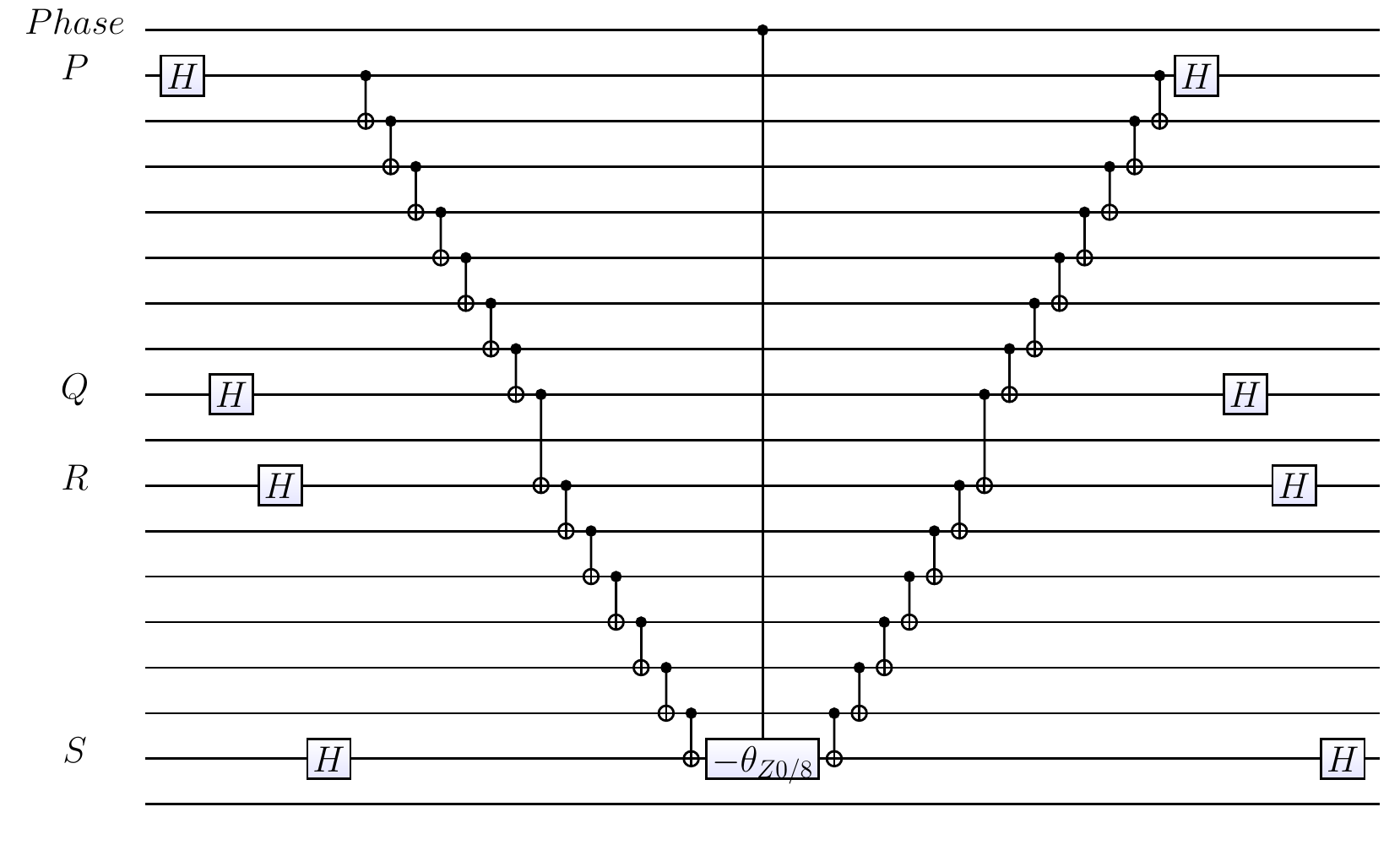}
 \caption{Standard circuit for $\Hpqrs$~\cite{whitfield2011}. $H$ represents Hadamard gate. Up to $8$ copies of this circuit appear in succession, with Hadamard gates replaced by other basis change gates to execute different subterms.\label{fig1} }
\end{figure}

Since the $\Hpqrs$ terms are the most numerous and thus dominate the computation time, we focus on speeding up the implementation of these terms; while the techniques can be straightforwardly used to also speed up $\Hpq$ or $\Hprrq$ terms, we prefer to execute those terms in a different order for reasons explained later.
The standard circuit~\cite{whitfield2011} to implement this is shown in Figure~\ref{fig1}. The lines represent qubits, with the top line labelled ``Phase" being used to control the application of the term. We show only a single subterm here; the circuit will be replicated several times with different subterms.
Our notation in writing circuits is that two qubit gates with open circles denote CNOTs, with the open circle representing the target of the CNOT. Boxes containing an $H$ represent Hadamard gates, which transform perform a basis change between the eigenbasis of $\sigma^X$ and $\sigma^Z$ operators and thus effective interchange these operators, i.e. $H \sigma^Z H = \sigma^X$. $Y$ and $Y^\dagger$ perform the analogous operation between $\sigma^Y$ and $\sigma^Z$ operators, i.e. $Y^\dagger \sigma^Y Y = \sigma^Z$. We thus refer to these as basis changes, and they are given by
\be
H=\frac{1}{\sqrt{2}}\begin{pmatrix} 1 & 1 \\ 1 & -1 \end{pmatrix} \quad  \quad
Y=\frac{1}{\sqrt{2}}\begin{pmatrix} 1 & i \\ i & 1 \end{pmatrix}.
\ee

\begin{figure}[t]
 \includegraphics[width=4in]{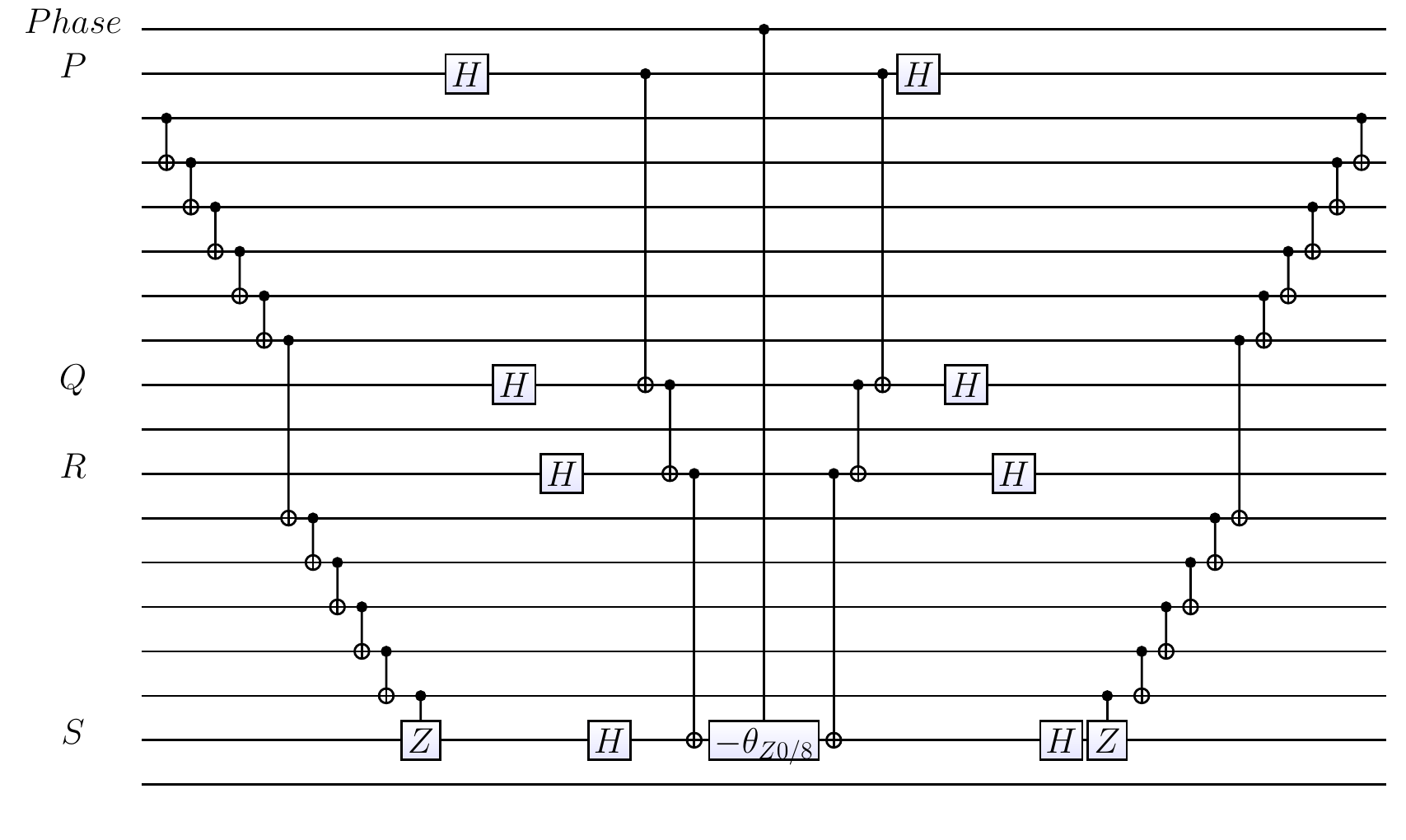}
 \caption{Alternative circuit for $\Hpqrs$. The $...$ indicate several different subterms. The two-qubit gate with a box labelled $Z$ is a controlled-$Z$ gate, executing $\sigma^Z$ on the target if the first qubit is $1$ and identity otherwise.} \label{fig1b}
\end{figure}

We notice immediately that the length of the Jordan-Wigner string may be proportional to $N$, leading to a linear overhead. However, as we now show, it is possible to replace this with a constant overhead.
Consider the circuit of Figure~\ref{fig1b}. The Jordan-Wigner string now appears outside all the basis changes, giving a constant factor improvement since it appears only once per term, rather than once per subterm.
The equivalence of the circuits Figure~\ref{fig1} and Figure~\ref{fig1b} can be worked out after some linear algebra. A brief explanation is as follows. Consider the subterm consisting of $\sigma^X_p \sigma^X_q \sigma^X_r \sigma^X_s$. If the product of the $Z$s appearing in the Jordan-Wigner string is equal to $+1$, then the controlled-Z has no effect and the gate implements the transformation $\exp(i \sigma^X_p \sigma^X_q \sigma^X_r \sigma^X_s \theta)$.
However, if the product is equal to $-1$, then the controlled-Z applies the operation $\sigma^Z_s$. This $\sigma^Z_s$ anti-commutes with $\sigma^X_s$, so that instead the gate implements $\exp(-i \sigma^X_p \sigma^X_q \sigma^X_r \sigma^X_s \theta)$, agreeing with Eq.~(\ref{pqrsunitary}).

\begin{figure}[t]
 \includegraphics[width=6in]{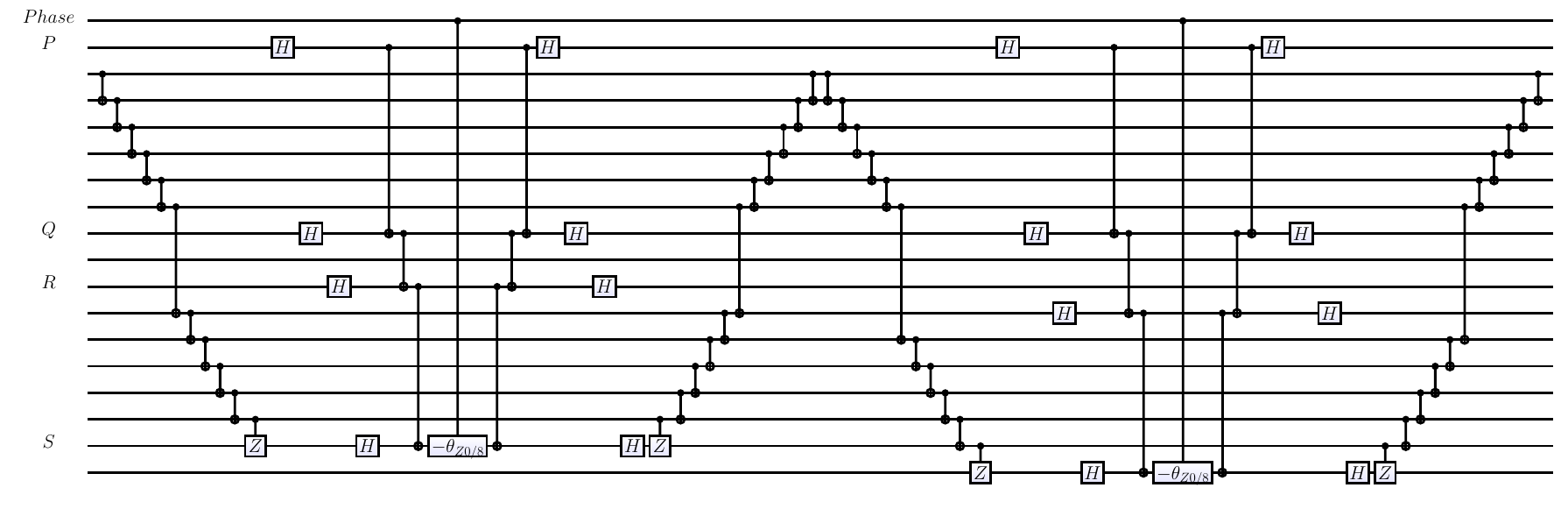}
 \caption{Alternative circuit for $\Hpqrs$ shown for two successive choices of $p,q,r,s$. In this case, both $r$ and $s$ changed so that some of the CNOT cannot be cancelled.} \label{multi}
\end{figure}

\begin{figure}[t]
 \includegraphics[width=6in]{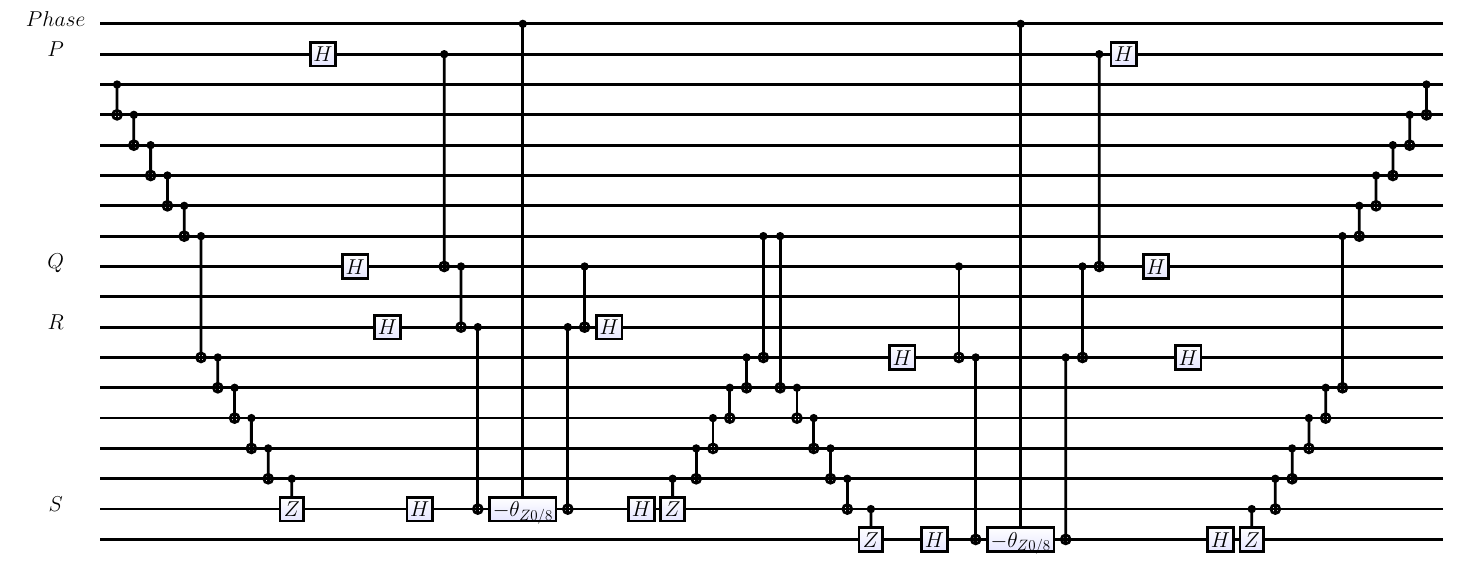}
 \caption{Equivalent circuit to Figure~\ref{multi}, with redundancies cancelled.} \label{multiredund}
\end{figure}

At this point, we have only achieved an improvement by a constant factor. However, a further improvement appears when we implement several of these terms in succession. For the Trotter-Suzuki decomposition, we have to implement every single possible term $\Hpqrs$. Let us order these terms lexicographically, meaning that we implement a given $\Hpqrs$ term, then increment $s$ and implement the next $\Hpqrs$ term; when $s=N$, we then increment $r$ and reduce $s$ to its minimum value, and so on.
In this case, the Jordan-Wigner strings largely cancel between successive terms; if we follow a term with given $p,q,r,s$ with a term with $p,q,r,s+1$, then all but $O(1)$ CNOTs can be cancelled between the end of one term and the start of the next, using the fact that the square of a CNOT gate is equal to the identity. If we follow the term with given $p,q,r,s$ with a term other than $p,q,r,s+1$, then more of the string of CNOTs may be left uncancelled; see Figures~\ref{multi},\ref{multiredund} for an illustration.

\begin{figure}[t]
 \includegraphics[width=6in]{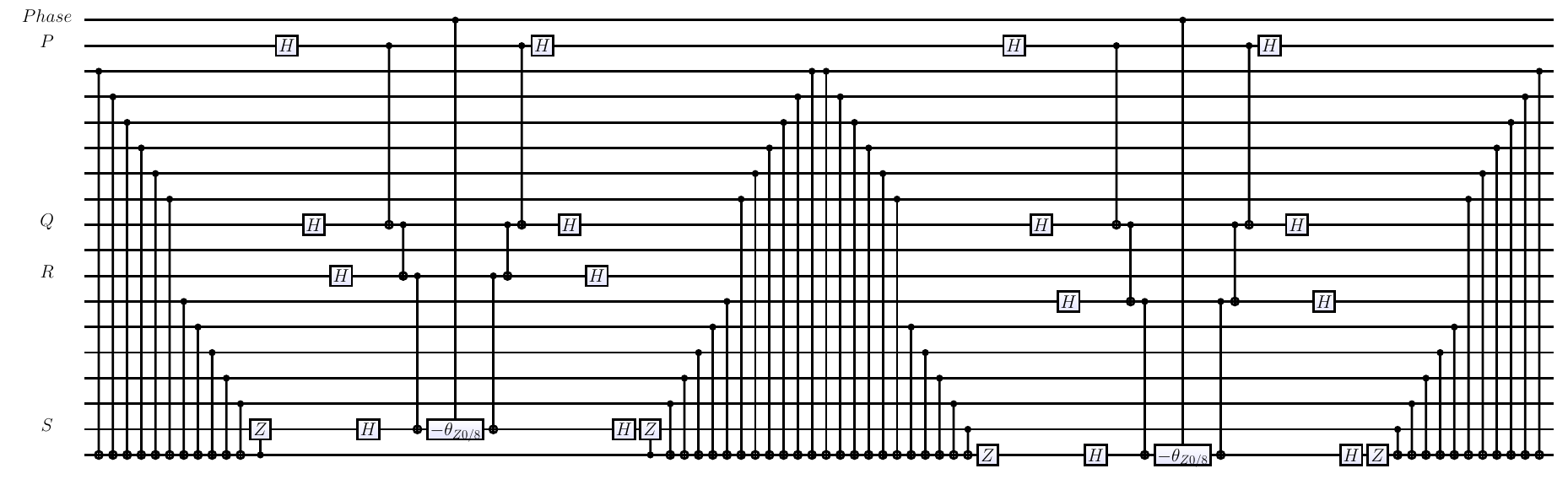}
 \caption{Circuit with ancilla for $\Hpqrs$ shown for two successive choices of $p,q,r,s$.} \label{multianc}
\end{figure}

There is a string of uncancelled CNOTs when we increment $r$ once $s=N$. However, since this occurs only every $O(N)$ steps, it leads to only subleading overhead. We can, however, cancel even these string by using the modified circuit shown in Figure~\ref{multianc} where we show several successive $\Hpqrs$ terms. Here we show a circuit that uses an additional ancilla (initialized to $|0\rangle$ at the start of the computation) to track the total parity of electrons on orbitals $p+1,...,q-1,r+1,...,s-1$. Then, it is possible to modify the lexicographic order by first increasing $s$ to $N$, then incrementing $r$ by $1$ and then decrementing $s$ from $N$, alternately incrementing and decrementing $s$ for each choice of $p,q,r$. This particular modification of the lexicographic order is {\it not} shown; rather we show a particular ordering that works well for parallelization as discussed later.

\begin{figure}[t]
 \includegraphics[width=6in]{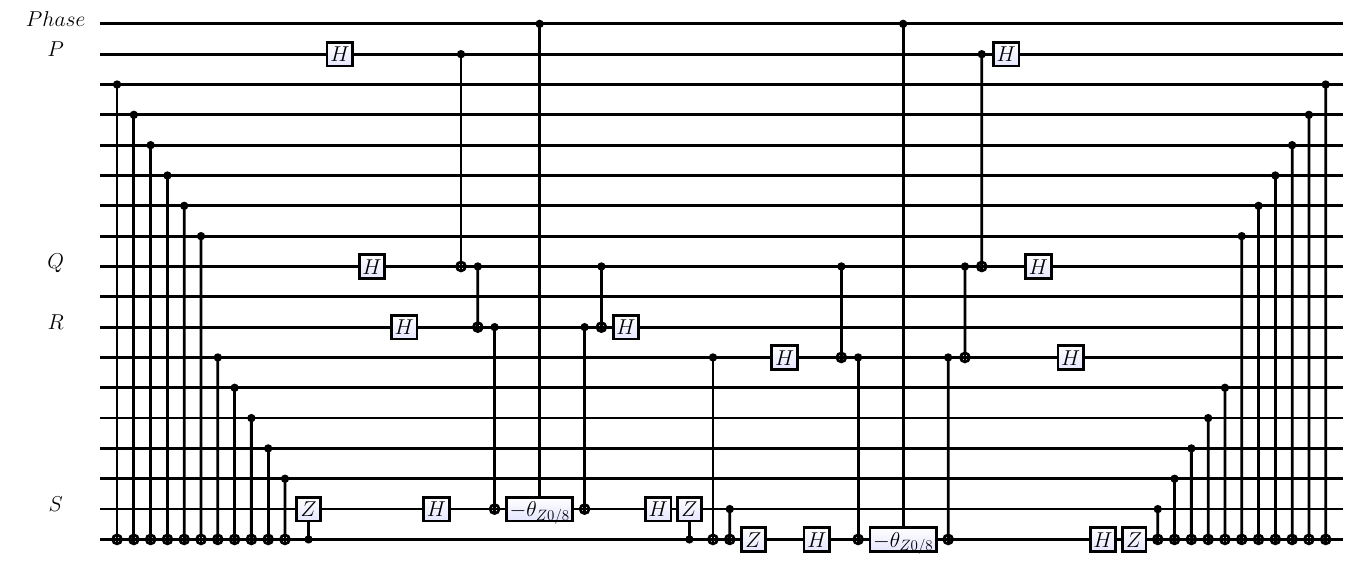}
 \caption{Equivalent circuit to Figure~\ref{multianc}, with redundancies cancelled. Compare to Figure~\ref{multiredund} where, without ancillas, not as much of the CNOT string could be cancelled.} \label{multiancredund}
\end{figure}

A further advantage of the ancilla circuit appears already without parallelization, in that it allows additional term cancellations. Note that for many molecules studied, not all $\Hpqrs$ terms are non-zero. Hence, it will often not be the case that given a particular $p,q,r,s$ that there is a non-zero term $p,q,r,s+1$. This at first seems to limit the number of cancellations as we often have uncancelled strings. However, in studying parallelization, we found that the following serendipitous cancellation often was possible: we often encountered a string of qubits acting with CNOTs on a given ancilla, with the qubits in that string in the order $b_1,b_2,...,b_n$, where $b_i$ labels a qubit. This string was then followed by some other string of qubits $c_1,c_2,...$. If $c_1=b_n, c_2=n_{n-1}$, and so on, then we can straightforwardly cancel successive CNOTs. However, note that since all the CNOTs in both strings have the same target ancilla, they all commute with each other, and hence we can cancel {\it any} CNOT in the first string against {\it any} CNOT in the second string, so long as they have the same source qubit (i.e., $b_i=c_j$ for some $i,j$). See Figure~\ref{multianc},\ref{multiancredund}.

The basic idea of all these techniques is simple: the Jordan-Wigner string computes the parity of a given set of qubits. If we then compute the parity of a set of qubits which is almost the same (for example, incrementing $s$ to $s+1$, adding a single qubit to that set), then much of the work of computation is already done and we do not need to do it again. In fact, similar improvements can be obtained using the original circuit of Figure~\ref{fig1} if, rather than choosing a specific choice of $p,q,r,s$ and then executing all needed subterms, we instead pick a given subterm and then execute all $p,q,r,s$ terms in lexicographic order using that subterm; however, this method does not allow the parallelization technique discussed next.

\section{Parallelization}
In previous work on simulation of quantum chemistry using quantum computers~\cite{qchem}, it was found that the main limitation
for practical applications will likely not be the number of coherent qubits available, but the number of gates that can be executed
coherently. Therefore, approaches that reduce the gate count at the cost of a few additional qubits may be highly useful.
We now introduce such an approach, which allows significant parallelization at the cost of using additional ancilla qubits.

Assume that we can execute gates in parallel that act on distinct qubits. There are many simple ways to parallelize the Trotter-Suzuki step.
For example, we can clearly execute two $\Hpqrs$ terms with $p,q,r,s$ and $p',q',r',s'$ at the same time if $p<q<r<s<p'<q'<r'<s'$, since the unitaries in the different gates act on different qubits. However, this can only lead to a constant factor parallelization improvement.

\begin{figure}
\label{nesting}
 \includegraphics[width=3.5in]{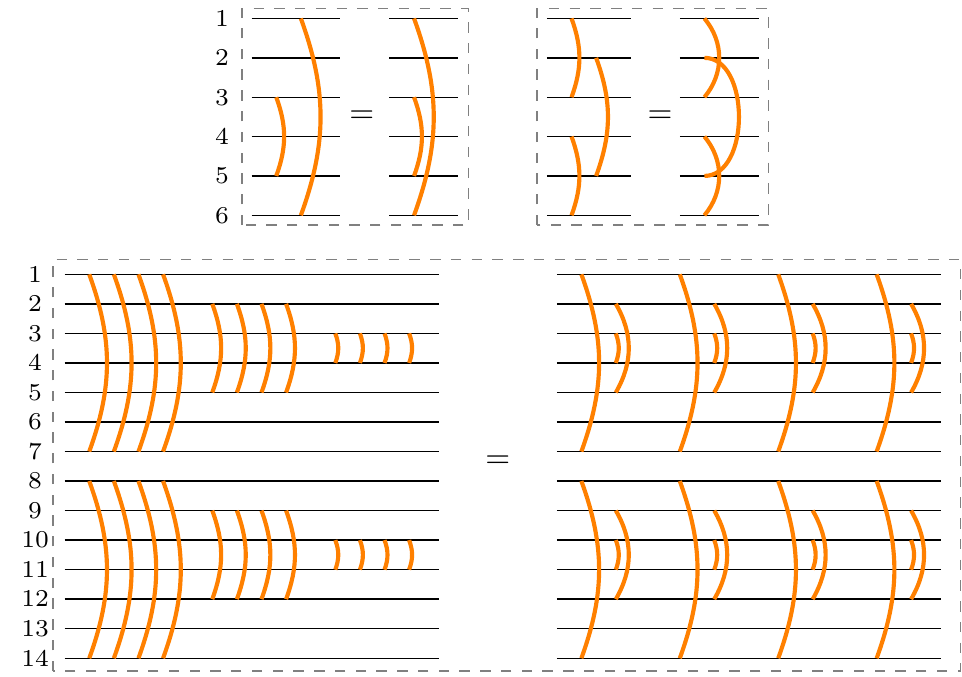}
 \caption{Schematic illustration of nesting procedure.
 {\it Top left panel:} Nesting the terms $h_{16}$ and $h_{35}$.
 {\it Top right panel:} Nesting the terms $h_{1436}$ and $h_{25}$. This is allowed because the lines cross an even number of times.
 {\it Bottom panel:} Nesting of three $h_{pqrs}$ terms, each with several subterms. This is a schematic representation of the example
 shown below in Figs.~\ref{threefull} and \ref{threenest}. }
\label{fignest}
\end{figure}

\begin{figure}[t]
 \includegraphics[width=5in]{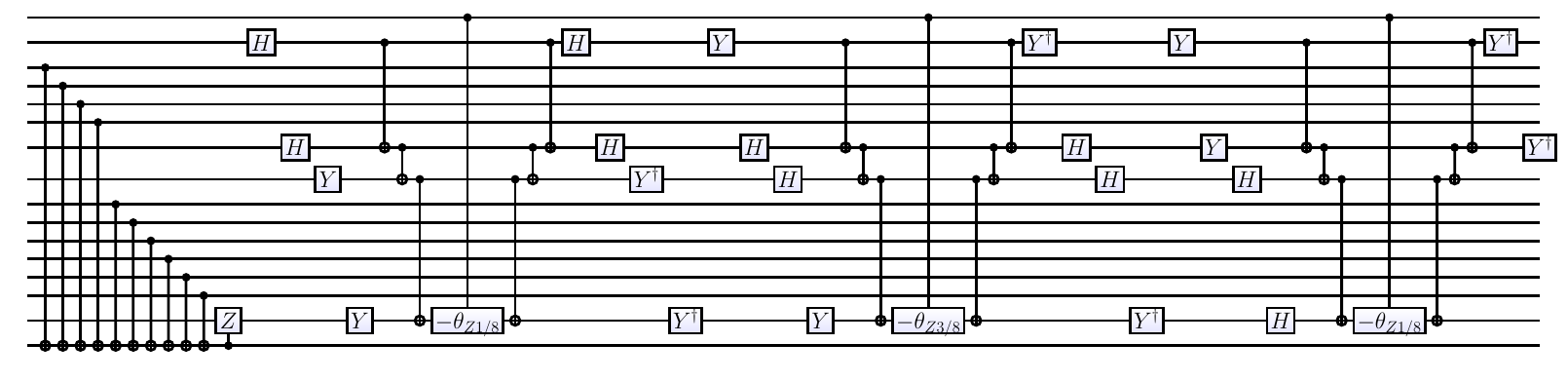}
 \includegraphics[width=5in]{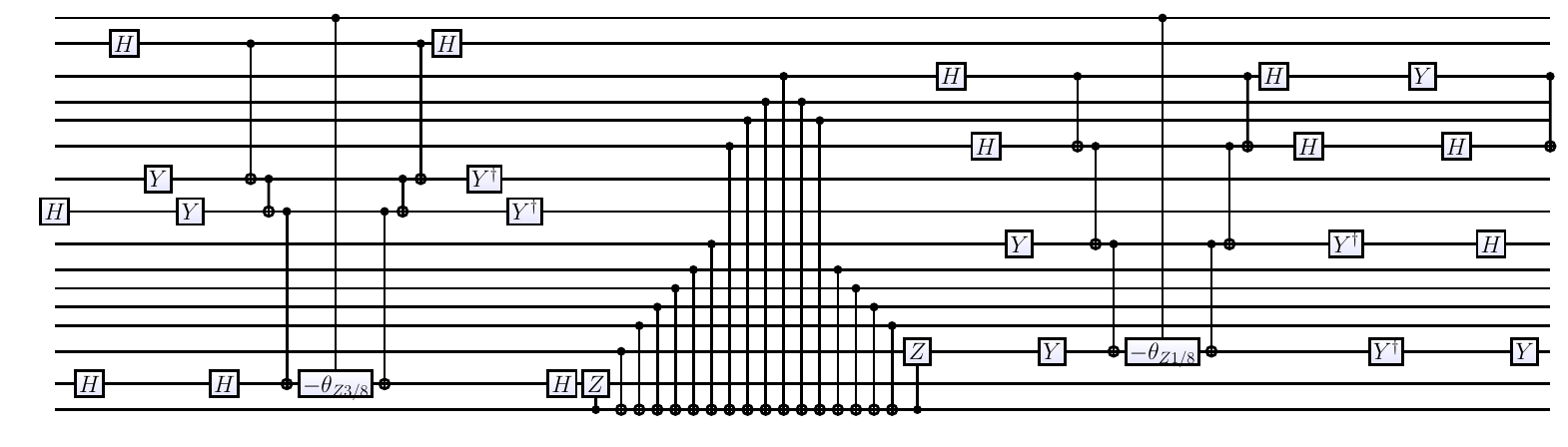}
 \includegraphics[width=5in]{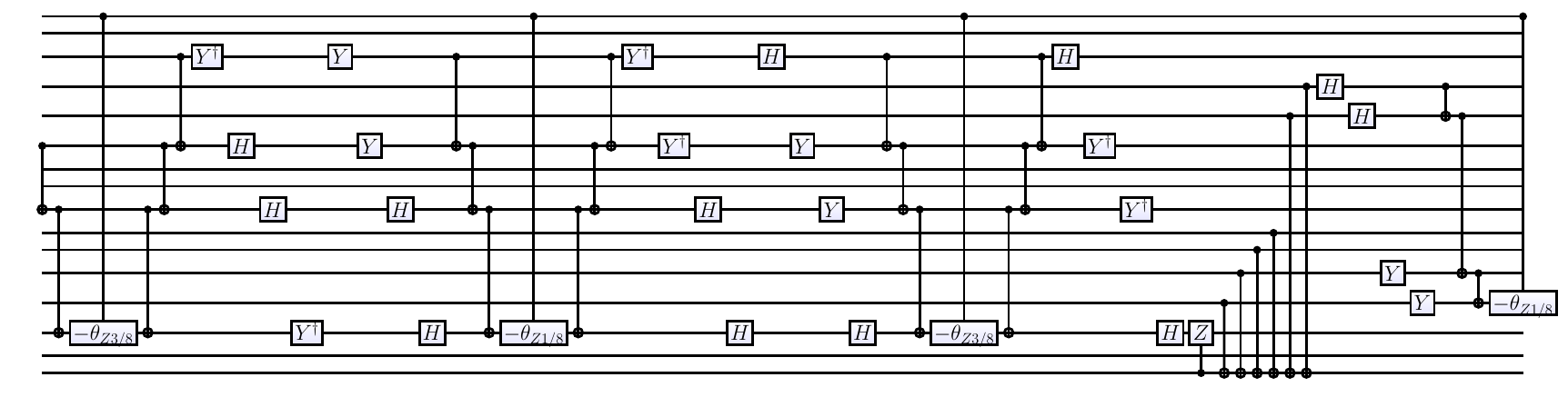}
 \includegraphics[width=5in]{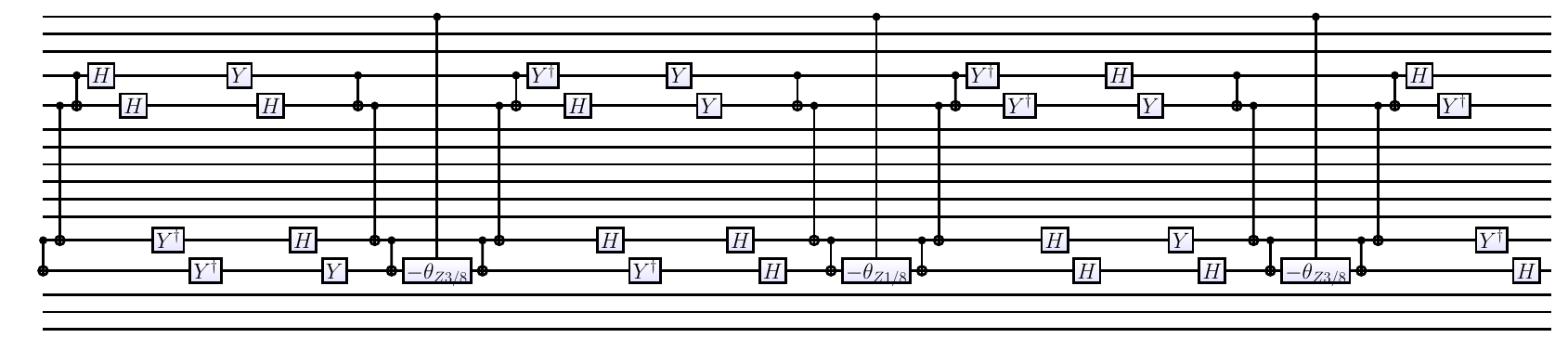}
 \caption{Example of a sequence of unnested $\Hpqrs$ terms.} \label{threefull}
\end{figure}

\begin{figure}[t]
 \includegraphics[width=5in]{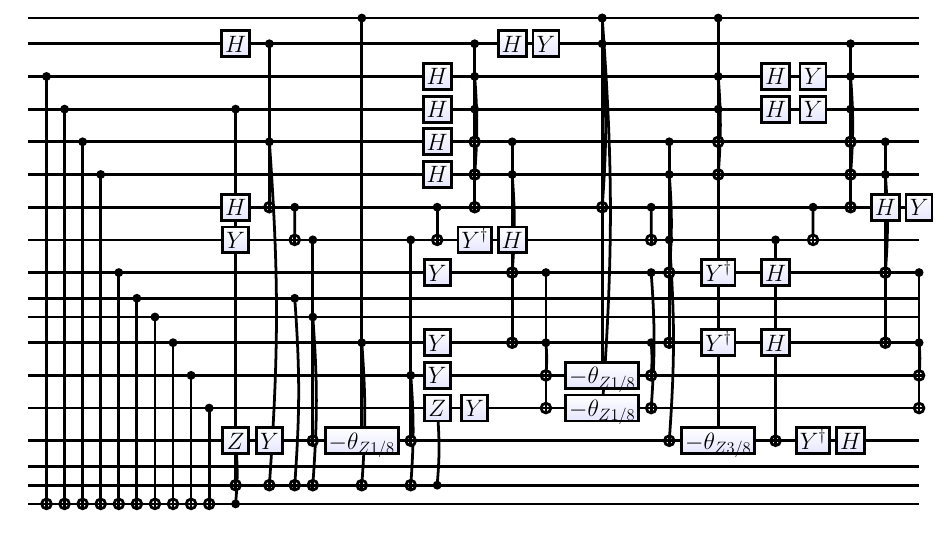}
 \includegraphics[width=5in]{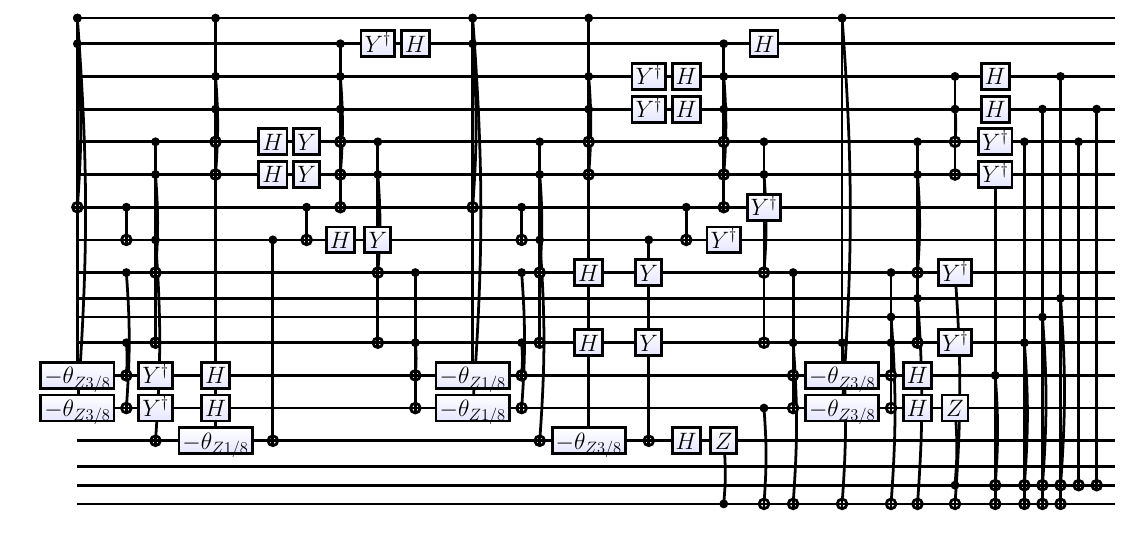}
 \caption{Circuit that implements the same unitary as Figure~\ref{threefull}, after nesting to reduce parallel depth.} \label{threenest}
\end{figure}

Using the circuit with additional ancilla qubits allows much better parallelization, improving parallel depth by a factor of $\Theta(N)$ compared to the serial gate count (thus giving a total $\Theta(N^2)$ improvement compared to the original).
Consider two $\Hpqrs$ terms. Given any choice of $p,q,r,s$ and $p',q',r',s'$ for which the sites $p',q',r',s'$ intersect an {\it even} number of sites in the Jordan-Wigner string of the $p,q,r,s$ term, the $p',q',r',s'$ term does not change the parity and hence can be moved through the Jordan-Wigner string. We refer to this as {\it nesting}, since we can execute terms in parallel when one sits inside another (for example, when $p<p'<q'<q$ and $r<r'<s'<s$), as illustrated in Figure~\ref{fignest}.
For a specific illustration,
Figure~\ref{threenest} shows the circuit of Figure~\ref{threefull} after nesting and CNOT cancellation; both of these equivalent circuits show a few terms in succession from a real molecule.
Different orderings are possible and we have not fully explored the optimum ones, but
up to $\Theta(N)$ terms can be executed in parallel this way with an appropriate term ordering.

\section{Implementation on \lqd}
To determine gate counts, the circuit for a single Jordan-Wigner step was specified by giving a sequence of $p,q,r,s$ and then the various circuits described above were used to implement each such term; then, \lqd automatically looked for cancellation of successive gate using rules such as the square of a CNOT or a Hadamard being equal to the identity and also automatically looked for possible nestings. \lqd also used the commutation rule discussed at the end of the above section to look for possible cancellations between CNOTs appearing in different orders.

\begin{figure}[t]
\includegraphics[width=\columnwidth]{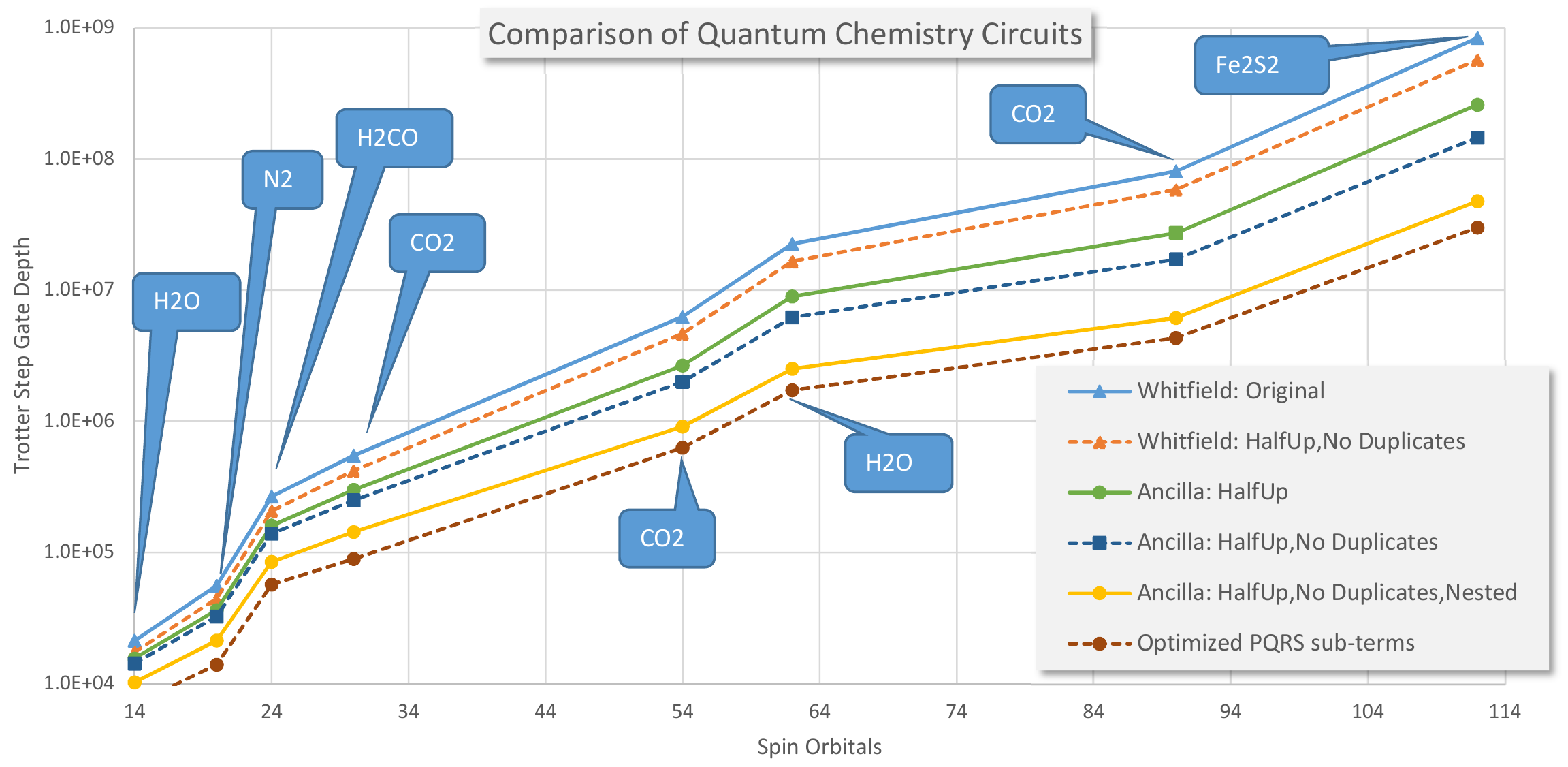}
 \caption{Gate counts for various circuits, as a function of number of spin orbitals, for various molecules. Horizontal axis indicates number of spin orbitals; each point is labelled by the corresponding molecule and certain molecules appear multiple times as the same molecule was studied with different basis sets. Vertical axis denotes gate depth. The lines marked Whitfield denote the original Whitfield et. al. circuit\cite{whitfield2011}. ``HalfUp" denotes a change in ordering of the qubits (see text). Ancilla denotes the use of the ancilla circuit. ``No Duplicates" means that \lqd searched for gates that could be cancelled, such as pairs of CNOTs that square to the identity. ``Nested" refers to the use of nesting. The bottom line, marked ``Optimized PQRS sub-terms" is ``Ancilla: HalfUp, No Duplicates, Nested" with a modified sequence of subterms, described in text.}
\label{fig3}
\end{figure}

The results are summarized in Figure~\ref{fig3} for a sequence of simple molecules, showing the improvement over the original circuit, as a function of the number of spins orbitals, for a number of real molecules. The $\Hpq$ and $\Hpqrs$ terms were obtained using PyQuante~\cite{pyquante}, an open source quantum chemistry program. While the number of $\Hpqrs$ terms scales as $N^4$, many of them are zero due to symmetries of these simple molecules; we expect that the gains would be even larger for larger and less-symmetric molecules when all $\Hpqrs$ terms were non-zero because when many of the terms are zero, the string cancellation is only partial.

For all cases of gate counts shown, the count shown is the parallel depth, where \lqd used the rule that it can execute gates in parallel which act on distinct qubits.  Further, we allowed the execution of gates in parallel even if more than one of them is controlled by the phase estimation ancilla, so long as all the other qubits in the gates are distinct; that is, we allow the phase estimation ancilla to control more than one rotation at a time.  To justify this last choice, suppose that we had {\it not} allowed the execution of such controlled rotations in parallel; we will now show that at the cost of a small number of additional ancilla qubits,
we could still obtain exactly the same parallel depth as in the case where we allow the phase estimation ancilla to control multiple rotations.
To do this, immediately before performing any of the Trotter-Suzuki evolution steps, we would copy (in the $S^z$ basis) the state of the phase estimation ancilla onto some number of additional ancilla qubits.  Then, we would use these additional ancillas to control the rotations so that every controlled rotation was controlled by a distinct qubit and the gates did not overlap.  Finally, at the end of all the Trotter-Suzuki evolution steps we would apply CNOT gates from the phase estimation ancilla onto the additional ancillas to restore them all to the $|0\rangle$ state, obtaining the desired unitary evolution.

The $\Theta(N)$ gains due to removing Jordan-Wigner strings and parallelization do lead to only modest improvement for these size molecules, but they are indeed seen to increase with $N$.  The $\Theta(N^2)$ improvement is not obvious in these figures, but we have found that the ratios in gate count (when going from the Whitfield et. al. circuit to the ancillas, or when adding nesting to the ancillas) can be approximately fit to a linear function of $N$, $a+b*N$ with $b<<a$, suggesting that at larger $N$ the relative gains will indeed become more significant.

At the same time, some seemingly minor improvements led to a large constant factor improvement in gate depth. One was to change the ordering of the qubits, from the order of Ref.~\onlinecite{qchem} where the qubits corresponded to (in order) the up and down spin states on the first orbital, then the up and down spin states and the second orbital, and so on, to a different order where first all up spin states were encoded and then all down spin states; this new order is referred to as ``HalfUp" in Figure~\ref{fig3}.
Another improvement is to change the sequence of subterms from that in Ref.~\onlinecite{whitfield2011} to a sequence which {\it minimized} the number of ways in which the basis changed at each step. Thus, rather than executing in the order $$\sigma^X \sigma^X \sigma^X \sigma^X,\sigma^Y \sigma^Y \sigma^Y \sigma^Y,\sigma^X \sigma^X \sigma^Y \sigma^Y,\sigma^Y \sigma^Y \sigma^X \sigma^X$$ we change to $$\sigma^X \sigma^X \sigma^X \sigma^X,\sigma^X \sigma^X \sigma^Y \sigma^Y,\sigma^Y \sigma^Y \sigma^Y \sigma^Y,\sigma^Y \sigma^Y \sigma^X \sigma^X,$$ allowing further cancellation of the basis change gates since two basis changes cancel at each step. This modified sequence is referred to as ``Optimized PQRS sub-terms" in Figure~\ref{fig3}.

All simulations were implemented using a first order Trotter-Suzuki decomposition. Similar improvements occur using a second order Trotter-Suzuki decomposition, but in fact in this case the first order Trotter-Suzuki gives the same error scaling as a second order one, as shown in App.~\ref{sct:appendix}.

\section{Trotter-Suzuki Term Ordering}
There are many possible orderings of the terms in the Trotter-Suzuki decomposition (lexicographic, by magnitude, etc...) and certain term orderings can greatly reduce the error at a given Trotter step $\delta_t$.
In the standard Hartree-Fock approach to a quantum chemistry problem, a variational state is constructed by finding a basis of single-particle states and then occupying certain orbitals, while others are empty. A variational optimization is then performed over the single-particle states and occupation vectors.
Let $n_r=0,1$ be the occupancy of the $r$-th orbital in the lowest-energy solution.
Then, in the Hartree-Fock basis for all $p \neq q$, \footnote{To extremize the Hartree-Fock energy, Eq.~(\ref{HFmin}) must hold for $p$ occupied and $q$ unoccupied. A further basis rotation which does not mix occupied and unoccupied orbitals makes it hold for all $p \neq q$}
\be
\label{HFmin}
t_{pq}+\frac{1}{2}\sum_r V_{prrq} n_r=0.
\ee
That is, the ``effective" hopping ($t_{pq}$ corrected by interaction terms) becomes diagonal in this basis.

One finds that for many molecules, the Hartree-Fock solution is a reasonable approximation, with the occupancies of the Hartree-Fock orbitals in the full ground state obtained by FCI being close to $0$ or $1$. Hence, the above equation holds to reasonable approximation in expectation. This suggests the following ``Interleaved Term Order":
\begin{itemize}
\item[{\bf 1.}] Execute all $\Hpp$ and $\Hpqqp$ terms.

\item[{\bf 2.}] For each $p,q$ {\bf do}
\begin{itemize}
\item[{\bf a.}] Do $\Hpq$ and all $\Hprrq$ terms.
\end{itemize}

\item[{\bf 3.}] Do all $\Hpqrs$ terms.
\end{itemize}
The terms in {\bf 1} commute with each other, as do those in {\bf 2a}, and thus can be executed in any order. Since the terms in {\bf 3} are the most numerous by far, we execute them in an order to minimize parallel depth as described above.

\begin{figure}[t]
 \includegraphics[width=\columnwidth]{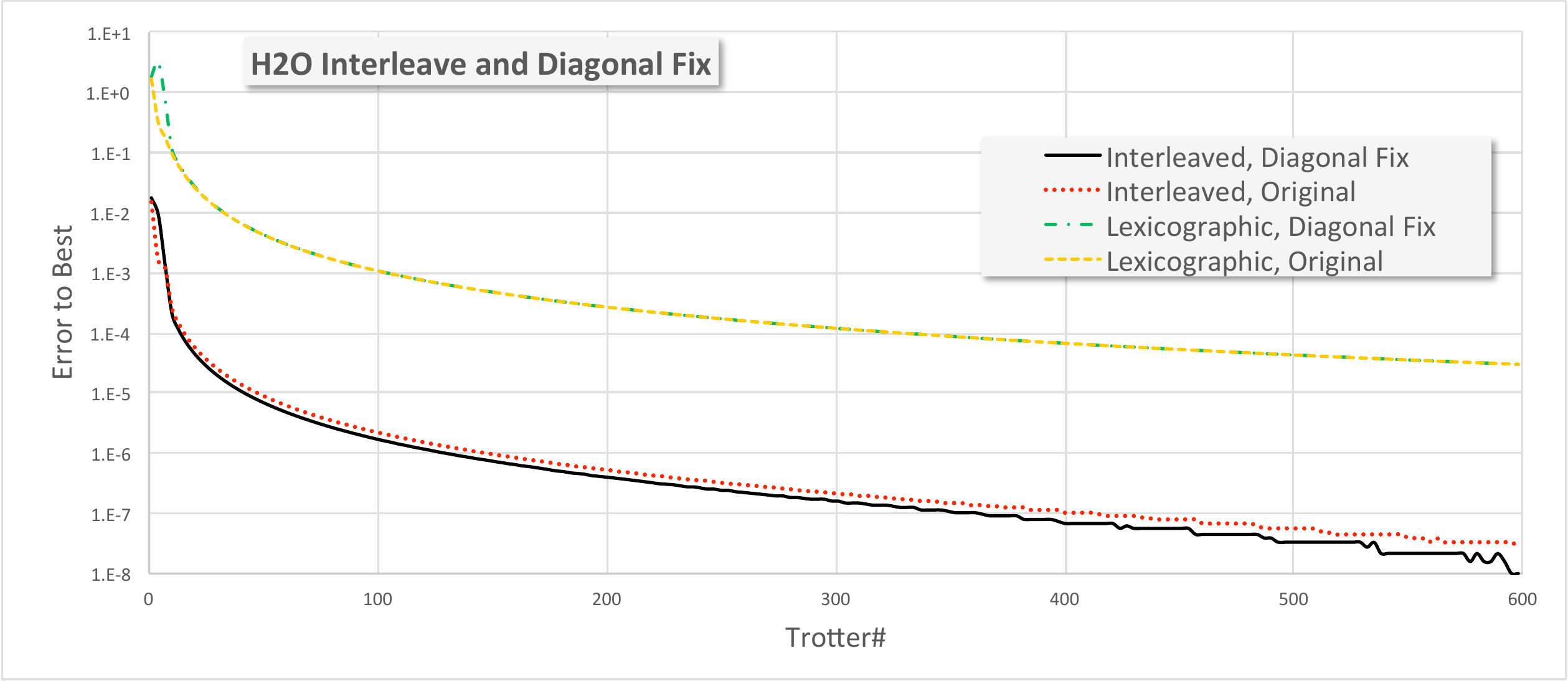}
 \caption{Trotter error for H$_2$O, as a function of Trotter number $n$, for time step $\delta_t=1/n$. Lexicographic (Lex) and Interleaved (Int) orders; Fix is explained in section \ref{corrH}. $y$-axis is base-10 logarithm of error in ground state measured in Hartrees.}
\label{fig4}
\end{figure}

Unfortunately, we do not yet have a theory that allows us to understand the scaling of the error for large molecules, but we empirically find consistent
large reductions in Trotter error using this ordering, largely because the $\Hpq$ terms are much larger than the others.
Figure~\ref{fig4} shows the reduction in Trotter error for H$_2$O. Similar though slightly smaller, improvements were seen for other molecules; one notable exception was for $N_2$, where the Hartree-Fock basis found by PyQuante poorly captured the occupied orbitals in the real molecule. We expect that working in the correct single-particle basis would lead to a large gain there as well.

In general, this approach relies on working in a basis in which Eq.~(\ref{HFmin}) holds, and thus it is most applicable to quantum chemistry; for problems such as the Hubbard model, it is possible to find such a basis of course (using a basis of Fourier modes), but this leads to a large overhead by increasing the number of terms in the Hamiltonian. On a speculative note, it may be possible in the Hubbard model to use a wavelet basis which would preserve approximate locality while still approximately satisfying Eq.~(\ref{HFmin}).

\section{Corrected Hamiltonian}
\label{corrH}
The final improvement leads to only a minor improvement (see Figure~\ref{fig4}), but the idea may be useful more generally. For a given $\delta_t$, rather than simulating Hamiltonian $H$, we simulate a different Hamiltonian, in which the coefficients $t_{pq}$ and $V_{pqrs}$ are modified in a way that goes to zero as $\delta_t\rightarrow0$.

To define the modification, we separate the Hamiltonian into diagonal ($\Hpp$ and $\Hpqqp$) terms, and off-diagonal terms. We make an approximation in the spirit of Hartree-Fock, defining an effective diagonal term $\omega_p=t_{pp}+(1/2) \sum_q V_{pqqp}$. We then calculate for a Hamiltonian $H=H_{\rm diag}+H_\text{off-diag}$ the effect of the off-diagonal term to second order in perturbation theory. The shift in energy $E_i$ of eigenstate $i$ is given by
$\sum_{j \neq i} (E_i-E_j)^{-1} |\langle \psi_j | H_{\rm off-diag} |\psi_j \rangle|^2$. However, using a second-order Trotter-Suzuki expression, the
energy shift is instead $\sum_{j \neq i} (E_i-E_j)^{-1} f(\delta_t(E_i-E_j))^{-2} |\langle \psi_j | H_\text{off-diag} |\psi_j \rangle|^2$, where
\be
f(x)=\sqrt{\frac{2 (1-\cos(x))}{\sin(x)}}.
\ee
To compensate, we make the modification
\be
V_{pqrs}\leftarrow V_{pqrs} f(\delta_t (\omega_p + \omega_q - \omega_r -\omega_s)),
\ee
and $t_{pq} \leftarrow t_{pq} f(\delta_t (\omega_p-\omega_q))$.

While only justified to second order, we find that this expression gives a slight consistent improvement in all cases. See the lines labelled ``Diagonal Fix" in Figure \ref{fig4}.
The function $f$ is slightly larger than $1$ for $x\neq 0$, so it increases the off-diagonal terms and goes to $1$ as $\delta_t \rightarrow 0$. Heuristically, this can be justified as follows: for large $E_i-E_j$, the effect of the off-diagonal term should produce only a virtual transition lasting time $\sim (E_i-E_j)^{-1}$; however, the minimum time of the transition is $\delta_t$, removing certain short-time transitions; hence we increase the magnitude of the term to compensate.

This modification can be thought of as using a ``renormalization group corrected Hamiltonian". Perhaps a more general approach can allow the addition of other terms to the Hamiltonian to compensate; however, we would like to avoid adding any terms involving six or more Fermi operators as they would be computationally expensive to implement.

\section{Discussion}
We have described several improvements to the basic Trotter-Suzuki approach used in the quantum simulation of quantum chemistry.  The cancellation of Jordan-Wigner strings and the nesting improve the $N^9$ scaling of the gate depth found in Ref.~\onlinecite{qchem} to $O(N^7)$.  Using approaches based on teleportation, it is possible to execute a sequence of measurements in parallel to implement in time $O(1)$ a CNOT string\cite{CodyJones2012} and even to implement, in certain coding schemes, an arbitrary Clifford operation\cite{Fowler2012}; with these approaches the gate depth using previous circuits scales as $N^8$.  Thus, even if all Clifford operations are assumed to be free (i.e., the cost of a Clifford operation is negligible compared to the cost of an arbitrary rotation), it is still worth using the circuits here as they enable additional parallelization of the rotations using nesting, reducing the time from $N^8$ scaling to $N^7$ scaling.

Assuming that the improvement from the interleaved term ordering continues to hold for larger molecules, we obtain a combined reduction in gate depth of two to three orders of magnitude for molecules such as Fe$_2$S$_2$ in a basis with $112$ spin orbitals, with larger improvements arising if we insist upon smaller Trotter error (see Figure \ref{fig4}).
Other schemes other than Trotter-Suzuki are certainly possible, with the two schemes with best theoretical scaling being Refs.~\onlinecite{berry2012black,berry2013exponential2}. However, Trotter-Suzuki has advantages of allowing improved parallelism, especially when each term commutes with many of the other terms. Also, the possibility of choosing different orderings allows us to exploit the ability to choose a Hartree-Fock basis in which the errors due to larger terms partly cancel. Future work using \lqd will allow accurate gate count comparisons using other schemes.

{\it Acknowledgments---} We thank B. K. Clark for useful discussions and collaboration on previous work.

\appendix
\section{Error Scaling in First Order Trotter-Suzuki} \label{sct:appendix}
Given a Hamiltonian $H$ written as a sum of terms
\be
H=\sum_{j=1}^{N_\text{terms}} H_j,
\ee
and given the ability to construct small quantum circuits to implement $\exp(i H_j \delta_t)$ for any $j$ and any $\delta_t$, the standard first order Trotter-Suzuki approximation for $\exp(i H \delta_t)$ is
\be
\exp(i H \delta_t) \approx \exp(i H_1 \delta_t) \exp(i H_2 \delta_t) ... \exp(i H_{N_\text{terms}} \delta_t),
\ee
while the second order Trotter-Suzuki approximation is
\begin{eqnarray}
&& \exp(i H \delta_t)\\ \nonumber
 &\approx & \exp\left(i H_1 \frac{\delta_t}{2}\right) \exp\left(i H_2 \frac{\delta_t}{2}\right) ... \exp\left(i H_{N_\text{terms}} \frac{\delta_t}{2}\right) \\ \nonumber &&
 \exp\left(i H_{N_{terms}} \frac{\delta_t}{2}\right) ...
 \exp\left(i H_2 \frac{\delta_t}{2}\right)
 \exp\left(i H_1 \frac{\delta_t}{2}\right).
\end{eqnarray}
As one may guess from the names, the first expression is exact to order $\delta_t$, giving an error of order $\delta_t^2$, while the second is exact to order $\delta_t^2$, giving an error of order $\delta_t^3$.

Ultimately, the goal is to approximate $\exp(i H t)$ for some large $t$. One does this by writing $\exp(i H t) = \exp(i H t/n)^n$ for some integer $n\geq 1$, and then using a Trotter-Suzuki approximation for $\exp(i H \delta_t)$ with $\delta_t=t/n$.
Taking into account that the we have decomposed the evolution for time $t$ into $n$ smaller steps, and using the error estimates for each time step above, one would expect that the error from first order Trotter-Suzuki scales as $1/n$ and that from second order Trotter-Suzuki scales as $1/n^2$.
In fact, in practice we have observed that both give errors {\it in the ground state energy}
scaling proportional to $1/n^2$; that is, first order Trotter-Suzuki has {\it better} error scaling than expected.

In this appendix we explain why this occurs, showing that this better error scaling is to be expected assuming that the Hamiltonian is real and has a non-degenerate ground state, both of which are true for the Hamiltonians considered in quantum chemistry. If instead we had a non-zero magnetic field, the Hamiltonian would not be real, and such better error scaling would not be expected.

Before explaining this, we note that the result holds trivially if $N_\text{terms}=2$, as then the first order and second order approximations to the unitary $\exp(i H \delta_t)$ have the same set of eigenvalues (to verify this, conjugate the first order approximation by the unitary $\exp(i H_2 \frac{\delta_t}{2})$ and this gives the second order approximation). However, our result holds for arbitrary $N_{terms}$.

To show the result, an explicit computation using commutators gives
\begin{eqnarray}
&& \exp(i H_1 \delta_t) \exp(i H_2 \delta_t) ... \exp(i H_{N_{terms}} \delta_t)
\\ \nonumber
&=& \exp(i (H+\Delta) \delta_t+O(\delta_t)^3),
\end{eqnarray}
where
\be
\Delta=i\sum_{j<k} [H_j,H_k] \frac{\delta_t}{2}.
\ee
Hence, up to an error which is $O(\delta_t^3)$, the first order Trotter-Suzuki formula computes the evolution under a modified Hamiltonian, $H+\Delta$. Let $E_0$ be the ground state energy of $H$ and let $E'_0$ be the ground state energy of $H+\Delta$, so that the error in ground state energy using a first order Trotter-Suzuki formula is equal to $E'_0-E_0+O(1/n^2)$. Assuming $H$ indeed has a non-degenerate ground state, then $E'_0-E_0$ is given by first order perturbation theory, up to error $O(\delta_t^2)$. However, if $H$ is real and has a non-degenerate ground state, then the ground state expectation value of $\Delta$ vanishes. Hence, $E'_0-E_0$ is $O(\delta_t^2)$.
So, in this case, indeed first order Trotter-Suzuki gives the correct energy up to error $1/n^2$.
To see that the ground state expectation value of $\Delta$ vanishes, let $\psi_0$ be the ground state wavefunction, which is real, while $\Delta$ is pure imaginary. Hence, the expectation value $\langle \psi_0 | \Delta | \psi_0 \rangle$ is pure imaginary; however, this expectation value is a real number, being the expectation value of a Hermitian operator, hence it equals zero.

It is interesting to observe that the non-degenerate ground state is necessary. Consider a single qubit with $H_1=\sigma_x,H_2=\sigma_z,H_3=-\sigma_x,H_4=-\sigma_z$ so that $H=0$ and $E_0=0$. One may verify that first order Trotter-Suzuki gives an error proportional to $1/n$ in this case.

\narrowtext
\bibliographystyle{apsrev4-1}
\bibliography{jw}

\begin{thebibliography}{40}%
\makeatletter
\providecommand \@ifxundefined [1]{%
 \@ifx{#1\undefined}
}%
\providecommand \@ifnum [1]{%
 \ifnum #1\expandafter \@firstoftwo
 \else \expandafter \@secondoftwo
 \fi
}%
\providecommand \@ifx [1]{%
 \ifx #1\expandafter \@firstoftwo
 \else \expandafter \@secondoftwo
 \fi
}%
\providecommand \natexlab [1]{#1}%
\providecommand \enquote  [1]{``#1''}%
\providecommand \bibnamefont  [1]{#1}%
\providecommand \bibfnamefont [1]{#1}%
\providecommand \citenamefont [1]{#1}%
\providecommand \href@noop [0]{\@secondoftwo}%
\providecommand \href [0]{\begingroup \@sanitize@url \@href}%
\providecommand \@href[1]{\@@startlink{#1}\@@href}%
\providecommand \@@href[1]{\endgroup#1\@@endlink}%
\providecommand \@sanitize@url [0]{\catcode `\\12\catcode `\$12\catcode
  `\&12\catcode `\#12\catcode `\^12\catcode `\_12\catcode `\%12\relax}%
\providecommand \@@startlink[1]{}%
\providecommand \@@endlink[0]{}%
\providecommand \url  [0]{\begingroup\@sanitize@url \@url }%
\providecommand \@url [1]{\endgroup\@href {#1}{\urlprefix }}%
\providecommand \urlprefix  [0]{URL }%
\providecommand \Eprint [0]{\href }%
\providecommand \doibase [0]{http://dx.doi.org/}%
\providecommand \selectlanguage [0]{\@gobble}%
\providecommand \bibinfo  [0]{\@secondoftwo}%
\providecommand \bibfield  [0]{\@secondoftwo}%
\providecommand \translation [1]{[#1]}%
\providecommand \BibitemOpen [0]{}%
\providecommand \bibitemStop [0]{}%
\providecommand \bibitemNoStop [0]{.\EOS\space}%
\providecommand \EOS [0]{\spacefactor3000\relax}%
\providecommand \BibitemShut  [1]{\csname bibitem#1\endcsname}%
\let\auto@bib@innerbib\@empty
\bibitem [{\citenamefont {Feynman}(1982)}]{Feynman1982}%
  \BibitemOpen
  \bibfield  {author} {\bibinfo {author} {\bibfnamefont {R.~P.}\ \bibnamefont
  {Feynman}},\ }\href {\doibase 10.1007/BF02650179} {\bibfield  {journal}
  {\bibinfo  {journal} {International Journal of Theoretical Physics}\ }\textbf
  {\bibinfo {volume} {21}},\ \bibinfo {pages} {467} (\bibinfo {year}
  {1982})}\BibitemShut {NoStop}%
\bibitem [{\citenamefont {Jordan}\ \emph {et~al.}(2012)\citenamefont {Jordan},
  \citenamefont {Lee},\ and\ \citenamefont {Preskill}}]{qft}%
  \BibitemOpen
  \bibfield  {author} {\bibinfo {author} {\bibfnamefont {S.~P.}\ \bibnamefont
  {Jordan}}, \bibinfo {author} {\bibfnamefont {K.~S.~M.}\ \bibnamefont {Lee}},
  \ and\ \bibinfo {author} {\bibfnamefont {J.}~\bibnamefont {Preskill}},\
  }\href {\doibase 10.1126/science.1217069} {\bibfield  {journal} {\bibinfo
  {journal} {Science}\ }\textbf {\bibinfo {volume} {336}},\ \bibinfo {pages}
  {1130} (\bibinfo {year} {2012})}\BibitemShut {NoStop}%
\bibitem [{\citenamefont {Lidar}\ and\ \citenamefont {Wang}(1999)}]{lidar1999}%
  \BibitemOpen
  \bibfield  {author} {\bibinfo {author} {\bibfnamefont {D.~A.}\ \bibnamefont
  {Lidar}}\ and\ \bibinfo {author} {\bibfnamefont {H.}~\bibnamefont {Wang}},\
  }\href {\doibase 10.1103/PhysRevE.59.2429} {\bibfield  {journal} {\bibinfo
  {journal} {Phys. Rev. E}\ }\textbf {\bibinfo {volume} {59}},\ \bibinfo
  {pages} {2429} (\bibinfo {year} {1999})}\BibitemShut {NoStop}%
\bibitem [{\citenamefont {Ortiz}\ \emph {et~al.}(2001)\citenamefont {Ortiz},
  \citenamefont {Gubernatis}, \citenamefont {Knill},\ and\ \citenamefont
  {Laflamme}}]{ortiz2001}%
  \BibitemOpen
  \bibfield  {author} {\bibinfo {author} {\bibfnamefont {G.}~\bibnamefont
  {Ortiz}}, \bibinfo {author} {\bibfnamefont {J.~E.}\ \bibnamefont
  {Gubernatis}}, \bibinfo {author} {\bibfnamefont {E.}~\bibnamefont {Knill}}, \
  and\ \bibinfo {author} {\bibfnamefont {R.}~\bibnamefont {Laflamme}},\ }\href
  {\doibase 10.1103/PhysRevA.64.022319} {\bibfield  {journal} {\bibinfo
  {journal} {Phys. Rev. A}\ }\textbf {\bibinfo {volume} {64}},\ \bibinfo
  {pages} {022319} (\bibinfo {year} {2001})}\BibitemShut {NoStop}%
\bibitem [{\citenamefont {Aspuru-Guzik}\ \emph {et~al.}(2005)\citenamefont
  {Aspuru-Guzik}, \citenamefont {Dutoi}, \citenamefont {Love},\ and\
  \citenamefont {Head-Gordon}}]{AspuruGuzik2005}%
  \BibitemOpen
  \bibfield  {author} {\bibinfo {author} {\bibfnamefont {A.}~\bibnamefont
  {Aspuru-Guzik}}, \bibinfo {author} {\bibfnamefont {A.~D.}\ \bibnamefont
  {Dutoi}}, \bibinfo {author} {\bibfnamefont {P.~J.}\ \bibnamefont {Love}}, \
  and\ \bibinfo {author} {\bibfnamefont {M.}~\bibnamefont {Head-Gordon}},\
  }\href {\doibase 10.1126/science.1113479} {\bibfield  {journal} {\bibinfo
  {journal} {Science}\ }\textbf {\bibinfo {volume} {309}},\ \bibinfo {pages}
  {1704} (\bibinfo {year} {2005})}\BibitemShut {NoStop}%
\bibitem [{\citenamefont {Kassal}\ \emph {et~al.}(2008)\citenamefont {Kassal},
  \citenamefont {Jordan}, \citenamefont {Love}, \citenamefont {Mohseni},\ and\
  \citenamefont {Aspuru-Guzik}}]{kassal2008}%
  \BibitemOpen
  \bibfield  {author} {\bibinfo {author} {\bibfnamefont {I.}~\bibnamefont
  {Kassal}}, \bibinfo {author} {\bibfnamefont {S.~P.}\ \bibnamefont {Jordan}},
  \bibinfo {author} {\bibfnamefont {P.~J.}\ \bibnamefont {Love}}, \bibinfo
  {author} {\bibfnamefont {M.}~\bibnamefont {Mohseni}}, \ and\ \bibinfo
  {author} {\bibfnamefont {A.}~\bibnamefont {Aspuru-Guzik}},\ }\href {\doibase
  10.1073/pnas.0808245105} {\bibfield  {journal} {\bibinfo  {journal}
  {Proceedings of the National Academy of Sciences}\ }\textbf {\bibinfo
  {volume} {105}},\ \bibinfo {pages} {18681} (\bibinfo {year}
  {2008})}\BibitemShut {NoStop}%
\bibitem [{\citenamefont {Wang}\ \emph {et~al.}(2008)\citenamefont {Wang},
  \citenamefont {Kais}, \citenamefont {Aspuru-Guzik},\ and\ \citenamefont
  {Hoffmann}}]{wang2008}%
  \BibitemOpen
  \bibfield  {author} {\bibinfo {author} {\bibfnamefont {H.}~\bibnamefont
  {Wang}}, \bibinfo {author} {\bibfnamefont {S.}~\bibnamefont {Kais}}, \bibinfo
  {author} {\bibfnamefont {A.}~\bibnamefont {Aspuru-Guzik}}, \ and\ \bibinfo
  {author} {\bibfnamefont {M.~R.}\ \bibnamefont {Hoffmann}},\ }\href {\doibase
  10.1039/B804804E} {\bibfield  {journal} {\bibinfo  {journal} {Physical
  Chemistry Chemical Physics}\ }\textbf {\bibinfo {volume} {10}},\ \bibinfo
  {pages} {5388} (\bibinfo {year} {2008})}\BibitemShut {NoStop}%
\bibitem [{\citenamefont {Kassal}\ and\ \citenamefont
  {Aspuru-Guzik}(2009)}]{kassal2009}%
  \BibitemOpen
  \bibfield  {author} {\bibinfo {author} {\bibfnamefont {I.}~\bibnamefont
  {Kassal}}\ and\ \bibinfo {author} {\bibfnamefont {A.}~\bibnamefont
  {Aspuru-Guzik}},\ }\href {\doibase http://dx.doi.org/10.1063/1.3266959}
  {\bibfield  {journal} {\bibinfo  {journal} {The Journal of Chemical Physics}\
  }\textbf {\bibinfo {volume} {131}},\  (\bibinfo {year} {2009})}\BibitemShut
  {NoStop}%
\bibitem [{\citenamefont {Lanyon}\ \emph {et~al.}(2010)\citenamefont {Lanyon}
  \emph {et~al.}}]{Lanyon2010}%
  \BibitemOpen
  \bibfield  {author} {\bibinfo {author} {\bibfnamefont {B.~P.}\ \bibnamefont
  {Lanyon}} \emph {et~al.},\ }\href {\doibase 10.1038/nchem.483} {\bibfield
  {journal} {\bibinfo  {journal} {Nat Chem}\ }\textbf {\bibinfo {volume} {2}},\
  \bibinfo {pages} {106} (\bibinfo {year} {2010})}\BibitemShut {NoStop}%
\bibitem [{\citenamefont {Whitfield}\ \emph {et~al.}(2011)\citenamefont
  {Whitfield}, \citenamefont {Biamonte},\ and\ \citenamefont
  {Aspuru-Guzik}}]{whitfield2011}%
  \BibitemOpen
  \bibfield  {author} {\bibinfo {author} {\bibfnamefont {J.~D.}\ \bibnamefont
  {Whitfield}}, \bibinfo {author} {\bibfnamefont {J.}~\bibnamefont {Biamonte}},
  \ and\ \bibinfo {author} {\bibfnamefont {A.}~\bibnamefont {Aspuru-Guzik}},\
  }\href {\doibase 10.1080/00268976.2011.552441} {\bibfield  {journal}
  {\bibinfo  {journal} {Molecular Physics}\ }\textbf {\bibinfo {volume}
  {109}},\ \bibinfo {pages} {735} (\bibinfo {year} {2011})}\BibitemShut
  {NoStop}%
\bibitem [{\citenamefont {Kassal}\ \emph {et~al.}(2011)\citenamefont {Kassal},
  \citenamefont {Whitfield}, \citenamefont {Perdomo-Ortiz}, \citenamefont
  {Yung},\ and\ \citenamefont {Aspuru-Guzik}}]{kassal2011}%
  \BibitemOpen
  \bibfield  {author} {\bibinfo {author} {\bibfnamefont {I.}~\bibnamefont
  {Kassal}}, \bibinfo {author} {\bibfnamefont {J.~D.}\ \bibnamefont
  {Whitfield}}, \bibinfo {author} {\bibfnamefont {A.}~\bibnamefont
  {Perdomo-Ortiz}}, \bibinfo {author} {\bibfnamefont {M.-H.}\ \bibnamefont
  {Yung}}, \ and\ \bibinfo {author} {\bibfnamefont {A.}~\bibnamefont
  {Aspuru-Guzik}},\ }\href {\doibase 10.1146/annurev-physchem-032210-103512}
  {\bibfield  {journal} {\bibinfo  {journal} {Annual Review of Physical
  Chemistry}\ }\textbf {\bibinfo {volume} {62}},\ \bibinfo {pages} {185}
  (\bibinfo {year} {2011})}\BibitemShut {NoStop}%
\bibitem [{\citenamefont {Yung}\ \emph {et~al.}(2012)\citenamefont {Yung},
  \citenamefont {Whitfield}, \citenamefont {Boixo}, \citenamefont {Tempel},\
  and\ \citenamefont {Aspuru-Guzik}}]{yung2012introduction}%
  \BibitemOpen
  \bibfield  {author} {\bibinfo {author} {\bibfnamefont {M.-H.}\ \bibnamefont
  {Yung}}, \bibinfo {author} {\bibfnamefont {J.~D.}\ \bibnamefont {Whitfield}},
  \bibinfo {author} {\bibfnamefont {S.}~\bibnamefont {Boixo}}, \bibinfo
  {author} {\bibfnamefont {D.~G.}\ \bibnamefont {Tempel}}, \ and\ \bibinfo
  {author} {\bibfnamefont {A.}~\bibnamefont {Aspuru-Guzik}},\ }\href@noop {}
  {\bibfield  {journal} {\bibinfo  {journal} {Preprint}\ } (\bibinfo {year}
  {2012})},\ \Eprint {http://arxiv.org/abs/1203.1331} {arXiv:1203.1331}
  \BibitemShut {NoStop}%
\bibitem [{\citenamefont {Whitfield}(2013)}]{whitfield2013spin}%
  \BibitemOpen
  \bibfield  {author} {\bibinfo {author} {\bibfnamefont {J.~D.}\ \bibnamefont
  {Whitfield}},\ }\href@noop {} {\bibfield  {journal} {\bibinfo  {journal}
  {Preprint}\ } (\bibinfo {year} {2013})},\ \Eprint
  {http://arxiv.org/abs/1306.1147} {arXiv:1306.1147} \BibitemShut {NoStop}%
\bibitem [{\citenamefont {Yung}\ \emph {et~al.}(2013)\citenamefont {Yung},
  \citenamefont {Casanova}, \citenamefont {Mezzacapo}, \citenamefont {McClean},
  \citenamefont {Lamata}, \citenamefont {Aspuru-Guzik},\ and\ \citenamefont
  {Solano}}]{yung2013}%
  \BibitemOpen
  \bibfield  {author} {\bibinfo {author} {\bibfnamefont {M.-H.}\ \bibnamefont
  {Yung}}, \bibinfo {author} {\bibfnamefont {J.}~\bibnamefont {Casanova}},
  \bibinfo {author} {\bibfnamefont {A.}~\bibnamefont {Mezzacapo}}, \bibinfo
  {author} {\bibfnamefont {J.}~\bibnamefont {McClean}}, \bibinfo {author}
  {\bibfnamefont {L.}~\bibnamefont {Lamata}}, \bibinfo {author} {\bibfnamefont
  {A.}~\bibnamefont {Aspuru-Guzik}}, \ and\ \bibinfo {author} {\bibfnamefont
  {E.}~\bibnamefont {Solano}},\ }\href@noop {} {\bibfield  {journal} {\bibinfo
  {journal} {Preprint}\ } (\bibinfo {year} {2013})},\ \Eprint
  {http://arxiv.org/abs/1307.4326} {arXiv:1307.4326} \BibitemShut {NoStop}%
\bibitem [{\citenamefont {Lamata}\ \emph {et~al.}(2013)\citenamefont {Lamata},
  \citenamefont {Mezzacapo}, \citenamefont {Casanova},\ and\ \citenamefont
  {Solano}}]{lamata2013}%
  \BibitemOpen
  \bibfield  {author} {\bibinfo {author} {\bibfnamefont {L.}~\bibnamefont
  {Lamata}}, \bibinfo {author} {\bibfnamefont {A.}~\bibnamefont {Mezzacapo}},
  \bibinfo {author} {\bibfnamefont {J.}~\bibnamefont {Casanova}}, \ and\
  \bibinfo {author} {\bibfnamefont {E.}~\bibnamefont {Solano}},\ }\href@noop {}
  {\bibfield  {journal} {\bibinfo  {journal} {Preprint}\ } (\bibinfo {year}
  {2013})},\ \Eprint {http://arxiv.org/abs/1312.2849} {arXiv:1312.2849}
  \BibitemShut {NoStop}%
\bibitem [{\citenamefont {{Toloui}}\ and\ \citenamefont
  {{Love}}(2013)}]{toloui2013}%
  \BibitemOpen
  \bibfield  {author} {\bibinfo {author} {\bibfnamefont {B.}~\bibnamefont
  {{Toloui}}}\ and\ \bibinfo {author} {\bibfnamefont {P.~J.}\ \bibnamefont
  {{Love}}},\ }\href@noop {} {\bibfield  {journal} {\bibinfo  {journal}
  {Preprint}\ } (\bibinfo {year} {2013})},\ \Eprint
  {http://arxiv.org/abs/1312.2579} {arXiv:1312.2579} \BibitemShut {NoStop}%
\bibitem [{\citenamefont {Wecker}\ \emph {et~al.}()\citenamefont {Wecker},
  \citenamefont {Bauer}, \citenamefont {Clark}, \citenamefont {Hastings},\ and\
  \citenamefont {Troyer}}]{qchem}%
  \BibitemOpen
  \bibfield  {author} {\bibinfo {author} {\bibfnamefont {D.}~\bibnamefont
  {Wecker}}, \bibinfo {author} {\bibfnamefont {B.}~\bibnamefont {Bauer}},
  \bibinfo {author} {\bibfnamefont {B.~K.}\ \bibnamefont {Clark}}, \bibinfo
  {author} {\bibfnamefont {M.~B.}\ \bibnamefont {Hastings}}, \ and\ \bibinfo
  {author} {\bibfnamefont {M.}~\bibnamefont {Troyer}},\ }\href@noop {} {\
  }\Eprint {http://arxiv.org/abs/1312.1695} {arXiv:1312.1695} \BibitemShut
  {NoStop}%
\bibitem [{\citenamefont {Nakano}\ and\ \citenamefont
  {Sakai}(2011)}]{nakano2011}%
  \BibitemOpen
  \bibfield  {author} {\bibinfo {author} {\bibfnamefont {H.}~\bibnamefont
  {Nakano}}\ and\ \bibinfo {author} {\bibfnamefont {T.}~\bibnamefont {Sakai}},\
  }\href {\doibase 10.1143/JPSJ.80.053704} {\bibfield  {journal} {\bibinfo
  {journal} {J. Phys. Soc. Jpn.}\ }\textbf {\bibinfo {volume} {80}},\ \bibinfo
  {pages} {053704} (\bibinfo {year} {2011})}\BibitemShut {NoStop}%
\bibitem [{\citenamefont {L{\"a}uchli}\ \emph {et~al.}(2011)\citenamefont
  {L{\"a}uchli}, \citenamefont {Sudan},\ and\ \citenamefont
  {S{\o}rensen}}]{lauchli2011ground}%
  \BibitemOpen
  \bibfield  {author} {\bibinfo {author} {\bibfnamefont {A.~M.}\ \bibnamefont
  {L{\"a}uchli}}, \bibinfo {author} {\bibfnamefont {J.}~\bibnamefont {Sudan}},
  \ and\ \bibinfo {author} {\bibfnamefont {E.~S.}\ \bibnamefont
  {S{\o}rensen}},\ }\href {\doibase 10.1103/PhysRevB.83.212401} {\bibfield
  {journal} {\bibinfo  {journal} {Physical Review B}\ }\textbf {\bibinfo
  {volume} {83}},\ \bibinfo {pages} {212401} (\bibinfo {year}
  {2011})}\BibitemShut {NoStop}%
\bibitem [{\citenamefont {Capponi}\ \emph {et~al.}(2013)\citenamefont
  {Capponi}, \citenamefont {Derzhko}, \citenamefont {Honecker}, \citenamefont
  {L{\"a}uchli},\ and\ \citenamefont {Richter}}]{capponi2013numerical}%
  \BibitemOpen
  \bibfield  {author} {\bibinfo {author} {\bibfnamefont {S.}~\bibnamefont
  {Capponi}}, \bibinfo {author} {\bibfnamefont {O.}~\bibnamefont {Derzhko}},
  \bibinfo {author} {\bibfnamefont {A.}~\bibnamefont {Honecker}}, \bibinfo
  {author} {\bibfnamefont {A.~M.}\ \bibnamefont {L{\"a}uchli}}, \ and\ \bibinfo
  {author} {\bibfnamefont {J.}~\bibnamefont {Richter}},\ }\href {\doibase
  10.1103/PhysRevB.88.144416} {\bibfield  {journal} {\bibinfo  {journal} {Phys.
  Rev. B}\ }\textbf {\bibinfo {volume} {88}},\ \bibinfo {pages} {144416}
  (\bibinfo {year} {2013})}\BibitemShut {NoStop}%
\bibitem [{\citenamefont {Gan}\ and\ \citenamefont
  {Harrison}(2005)}]{gan2005calibrating}%
  \BibitemOpen
  \bibfield  {author} {\bibinfo {author} {\bibfnamefont {Z.}~\bibnamefont
  {Gan}}\ and\ \bibinfo {author} {\bibfnamefont {R.~J.}\ \bibnamefont
  {Harrison}},\ }in\ \href@noop {} {\emph {\bibinfo {booktitle}
  {Supercomputing, 2005. Proceedings of the ACM/IEEE SC 2005 Conference}}}\
  (\bibinfo {organization} {IEEE},\ \bibinfo {year} {2005})\ pp.\ \bibinfo
  {pages} {22--22}\BibitemShut {NoStop}%
\bibitem [{\citenamefont {Kurashige}\ \emph {et~al.}(2013)\citenamefont
  {Kurashige}, \citenamefont {Chan},\ and\ \citenamefont
  {Yanai}}]{Kurashige2013}%
  \BibitemOpen
  \bibfield  {author} {\bibinfo {author} {\bibfnamefont {Y.}~\bibnamefont
  {Kurashige}}, \bibinfo {author} {\bibfnamefont {G.~K.-L.}\ \bibnamefont
  {Chan}}, \ and\ \bibinfo {author} {\bibfnamefont {T.}~\bibnamefont {Yanai}},\
  }\href {\doibase 10.1038/nchem.1677} {\bibfield  {journal} {\bibinfo
  {journal} {Nat Chem}\ }\textbf {\bibinfo {volume} {5}},\ \bibinfo {pages}
  {660} (\bibinfo {year} {2013})}\BibitemShut {NoStop}%
\bibitem [{\citenamefont {Berry}\ \emph {et~al.}(2007)\citenamefont {Berry},
  \citenamefont {Ahokas}, \citenamefont {Cleve},\ and\ \citenamefont
  {Sanders}}]{efficient2007}%
  \BibitemOpen
  \bibfield  {author} {\bibinfo {author} {\bibfnamefont {D.}~\bibnamefont
  {Berry}}, \bibinfo {author} {\bibfnamefont {G.}~\bibnamefont {Ahokas}},
  \bibinfo {author} {\bibfnamefont {R.}~\bibnamefont {Cleve}}, \ and\ \bibinfo
  {author} {\bibfnamefont {B.}~\bibnamefont {Sanders}},\ }\href@noop {}
  {\bibfield  {journal} {\bibinfo  {journal} {Commun. Math. Phys.}\ }\textbf
  {\bibinfo {volume} {270}},\ \bibinfo {pages} {359} (\bibinfo {year}
  {2007})}\BibitemShut {NoStop}%
\bibitem [{\citenamefont {Childs}\ and\ \citenamefont
  {Kothari}(2011)}]{childs2011simulating}%
  \BibitemOpen
  \bibfield  {author} {\bibinfo {author} {\bibfnamefont {A.~M.}\ \bibnamefont
  {Childs}}\ and\ \bibinfo {author} {\bibfnamefont {R.}~\bibnamefont
  {Kothari}},\ }in\ \href {\doibase 10.1007/978-3-642-18073-6_8} {\emph
  {\bibinfo {booktitle} {Theory of Quantum Computation, Communication, and
  Cryptography}}},\ \bibinfo {series} {Lecture Notes in Computer Science},
  Vol.\ \bibinfo {volume} {6519},\ \bibinfo {editor} {edited by\ \bibinfo
  {editor} {\bibfnamefont {W.}~\bibnamefont {Dam}}, \bibinfo {editor}
  {\bibfnamefont {V.}~\bibnamefont {Kendon}}, \ and\ \bibinfo {editor}
  {\bibfnamefont {S.}~\bibnamefont {Severini}}}\ (\bibinfo  {publisher}
  {Springer Berlin Heidelberg},\ \bibinfo {year} {2011})\ pp.\ \bibinfo {pages}
  {94--103}\BibitemShut {NoStop}%
\bibitem [{\citenamefont {Daskin}\ and\ \citenamefont
  {Kais}(2011)}]{anmer2011}%
  \BibitemOpen
  \bibfield  {author} {\bibinfo {author} {\bibfnamefont {A.}~\bibnamefont
  {Daskin}}\ and\ \bibinfo {author} {\bibfnamefont {S.}~\bibnamefont {Kais}},\
  }\href {\doibase 10.1063/1.3575402} {\bibfield  {journal} {\bibinfo
  {journal} {The Journal of Chemical Physics}\ }\textbf {\bibinfo {volume}
  {134}},\  (\bibinfo {year} {2011})}\BibitemShut {NoStop}%
\bibitem [{\citenamefont {Childs}\ and\ \citenamefont
  {Wiebe}(2012)}]{childs2012hamiltonian}%
  \BibitemOpen
  \bibfield  {author} {\bibinfo {author} {\bibfnamefont {A.~M.}\ \bibnamefont
  {Childs}}\ and\ \bibinfo {author} {\bibfnamefont {N.}~\bibnamefont {Wiebe}},\
  }\href@noop {} {\bibfield  {journal} {\bibinfo  {journal} {Quantum
  Information \& Computation}\ }\textbf {\bibinfo {volume} {12}},\ \bibinfo
  {pages} {901} (\bibinfo {year} {2012})},\ \Eprint
  {http://arxiv.org/abs/1202.5822} {arXiv:1202.5822} \BibitemShut {NoStop}%
\bibitem [{\citenamefont {Childs}\ and\ \citenamefont
  {Wiebe}(2013)}]{childs2012product}%
  \BibitemOpen
  \bibfield  {author} {\bibinfo {author} {\bibfnamefont {A.~M.}\ \bibnamefont
  {Childs}}\ and\ \bibinfo {author} {\bibfnamefont {N.}~\bibnamefont {Wiebe}},\
  }\href {\doibase 10.1063/1.4811386} {\bibfield  {journal} {\bibinfo
  {journal} {J. Math. Phys.}\ }\textbf {\bibinfo {volume} {54}},\ \bibinfo
  {pages} {062202} (\bibinfo {year} {2013})},\ \Eprint
  {http://arxiv.org/abs/1211.4945} {arXiv:1211.4945} \BibitemShut {NoStop}%
\bibitem [{\citenamefont {Berry}\ and\ \citenamefont
  {Childs}(2012)}]{berry2012black}%
  \BibitemOpen
  \bibfield  {author} {\bibinfo {author} {\bibfnamefont {D.~W.}\ \bibnamefont
  {Berry}}\ and\ \bibinfo {author} {\bibfnamefont {A.~M.}\ \bibnamefont
  {Childs}},\ }\href@noop {} {\bibfield  {journal} {\bibinfo  {journal}
  {Quantum Information \& Computation}\ }\textbf {\bibinfo {volume} {12}},\
  \bibinfo {pages} {29} (\bibinfo {year} {2012})},\ \Eprint
  {http://arxiv.org/abs/0910.4157} {arXiv:0910.4157} \BibitemShut {NoStop}%
\bibitem [{\citenamefont {Daskin}\ \emph {et~al.}(2012)\citenamefont {Daskin},
  \citenamefont {Grama}, \citenamefont {Kollias},\ and\ \citenamefont
  {Kais}}]{anmer2012}%
  \BibitemOpen
  \bibfield  {author} {\bibinfo {author} {\bibfnamefont {A.}~\bibnamefont
  {Daskin}}, \bibinfo {author} {\bibfnamefont {A.}~\bibnamefont {Grama}},
  \bibinfo {author} {\bibfnamefont {G.}~\bibnamefont {Kollias}}, \ and\
  \bibinfo {author} {\bibfnamefont {S.}~\bibnamefont {Kais}},\ }\href {\doibase
  10.1063/1.4772185} {\bibfield  {journal} {\bibinfo  {journal} {The Journal of
  Chemical Physics}\ }\textbf {\bibinfo {volume} {137}},\  (\bibinfo {year}
  {2012})}\BibitemShut {NoStop}%
\bibitem [{\citenamefont {Berry}\ \emph {et~al.}(2013)\citenamefont {Berry},
  \citenamefont {Childs}, \citenamefont {Cleve}, \citenamefont {Kothari},\ and\
  \citenamefont {Somma}}]{berry2013exponential2}%
  \BibitemOpen
  \bibfield  {author} {\bibinfo {author} {\bibfnamefont {D.~W.}\ \bibnamefont
  {Berry}}, \bibinfo {author} {\bibfnamefont {A.~M.}\ \bibnamefont {Childs}},
  \bibinfo {author} {\bibfnamefont {R.}~\bibnamefont {Cleve}}, \bibinfo
  {author} {\bibfnamefont {R.}~\bibnamefont {Kothari}}, \ and\ \bibinfo
  {author} {\bibfnamefont {R.~D.}\ \bibnamefont {Somma}},\ }\href@noop {}
  {\bibfield  {journal} {\bibinfo  {journal} {Preprint}\ } (\bibinfo {year}
  {2013})},\ \Eprint {http://arxiv.org/abs/1312.1414} {arXiv:1312.1414}
  \BibitemShut {NoStop}%
\bibitem [{\citenamefont {Babbush}\ \emph {et~al.}(2013)\citenamefont
  {Babbush}, \citenamefont {Love},\ and\ \citenamefont
  {Aspuru-Guzik}}]{babbush2013}%
  \BibitemOpen
  \bibfield  {author} {\bibinfo {author} {\bibfnamefont {R.}~\bibnamefont
  {Babbush}}, \bibinfo {author} {\bibfnamefont {P.~J.}\ \bibnamefont {Love}}, \
  and\ \bibinfo {author} {\bibfnamefont {A.}~\bibnamefont {Aspuru-Guzik}},\
  }\href@noop {} {\bibfield  {journal} {\bibinfo  {journal} {Preprint}\ }
  (\bibinfo {year} {2013})},\ \Eprint {http://arxiv.org/abs/1311.3967}
  {arXiv:1311.3967} \BibitemShut {NoStop}%
\bibitem [{\citenamefont {Peruzzo}\ \emph {et~al.}(2013)\citenamefont
  {Peruzzo}, \citenamefont {McClean}, \citenamefont {Shadbolt}, \citenamefont
  {Yung}, \citenamefont {Zhou}, \citenamefont {Love}, \citenamefont
  {Aspuru-Guzik},\ and\ \citenamefont {O'Brien}}]{peruzzo2013}%
  \BibitemOpen
  \bibfield  {author} {\bibinfo {author} {\bibfnamefont {A.}~\bibnamefont
  {Peruzzo}}, \bibinfo {author} {\bibfnamefont {J.}~\bibnamefont {McClean}},
  \bibinfo {author} {\bibfnamefont {P.}~\bibnamefont {Shadbolt}}, \bibinfo
  {author} {\bibfnamefont {M.-H.}\ \bibnamefont {Yung}}, \bibinfo {author}
  {\bibfnamefont {X.-Q.}\ \bibnamefont {Zhou}}, \bibinfo {author}
  {\bibfnamefont {P.~J.}\ \bibnamefont {Love}}, \bibinfo {author}
  {\bibfnamefont {A.}~\bibnamefont {Aspuru-Guzik}}, \ and\ \bibinfo {author}
  {\bibfnamefont {J.~L.}\ \bibnamefont {O'Brien}},\ }\href@noop {} {\bibfield
  {journal} {\bibinfo  {journal} {Preprint}\ } (\bibinfo {year} {2013})},\
  \Eprint {http://arxiv.org/abs/1304.3061} {arXiv:1304.3061} \BibitemShut
  {NoStop}%
\bibitem [{\citenamefont {Trotter}(1959)}]{trotter1959}%
  \BibitemOpen
  \bibfield  {author} {\bibinfo {author} {\bibfnamefont {H.~F.}\ \bibnamefont
  {Trotter}},\ }\href@noop {} {\bibfield  {journal} {\bibinfo  {journal} {Proc.
  Amer. Math. Soc.}\ }\textbf {\bibinfo {volume} {10}},\ \bibinfo {pages} {545}
  (\bibinfo {year} {1959})}\BibitemShut {NoStop}%
\bibitem [{\citenamefont {Suzuki}(1976)}]{suzuki1976}%
  \BibitemOpen
  \bibfield  {author} {\bibinfo {author} {\bibfnamefont {M.}~\bibnamefont
  {Suzuki}},\ }\href {\doibase 10.1007/BF01609348} {\bibfield  {journal}
  {\bibinfo  {journal} {Communications in Mathematical Physics}\ }\textbf
  {\bibinfo {volume} {51}},\ \bibinfo {pages} {183} (\bibinfo {year}
  {1976})}\BibitemShut {NoStop}%
\bibitem [{\citenamefont {Bravyi}\ and\ \citenamefont {Kitaev}(2002)}]{BK}%
  \BibitemOpen
  \bibfield  {author} {\bibinfo {author} {\bibfnamefont {S.}~\bibnamefont
  {Bravyi}}\ and\ \bibinfo {author} {\bibfnamefont {A.}~\bibnamefont
  {Kitaev}},\ }\href@noop {} {\bibfield  {journal} {\bibinfo  {journal} {Ann.
  Phys.}\ }\textbf {\bibinfo {volume} {298}},\ \bibinfo {pages} {210} (\bibinfo
  {year} {2002})}\BibitemShut {NoStop}%
\bibitem [{\citenamefont {Wecker}\ and\ \citenamefont
  {Svore}(2014)}]{wecker2014}%
  \BibitemOpen
  \bibfield  {author} {\bibinfo {author} {\bibfnamefont {D.}~\bibnamefont
  {Wecker}}\ and\ \bibinfo {author} {\bibfnamefont {K.}~\bibnamefont {Svore}},\
  }\href@noop {} {\bibfield  {journal} {\bibinfo  {journal} {Preprint}\ }
  (\bibinfo {year} {2014})},\ \Eprint {http://arxiv.org/abs/1402.4467}
  {arXiv:1402.4467} \BibitemShut {NoStop}%
\bibitem [{pyq()}]{pyquante}%
  \BibitemOpen
  \href@noop {} {}\bibinfo {note} {{Available at
  http://pyquante.sourceforge.net}}\BibitemShut {NoStop}%
\bibitem [{Note1()}]{Note1}%
  \BibitemOpen
  \bibinfo {note} {To extremize the Hartree-Fock energy, Eq.~(\ref {HFmin})
  must hold for $p$ occupied and $q$ unoccupied. A further basis rotation which
  does not mix occupied and unoccupied orbitals makes it hold for all $p \not
  =q$}\BibitemShut {NoStop}%
\bibitem [{\citenamefont {Cody~Jones}\ \emph {et~al.}(2012)\citenamefont
  {Cody~Jones}, \citenamefont {Whitfield}, \citenamefont {McMahon},
  \citenamefont {Yung}, \citenamefont {Meter}, \citenamefont {Aspuru-Guzik},\
  and\ \citenamefont {Yamamoto}}]{CodyJones2012}%
  \BibitemOpen
  \bibfield  {author} {\bibinfo {author} {\bibfnamefont {N.}~\bibnamefont
  {Cody~Jones}}, \bibinfo {author} {\bibfnamefont {J.~D.}\ \bibnamefont
  {Whitfield}}, \bibinfo {author} {\bibfnamefont {P.~L.}\ \bibnamefont
  {McMahon}}, \bibinfo {author} {\bibfnamefont {M.-H.}\ \bibnamefont {Yung}},
  \bibinfo {author} {\bibfnamefont {R.~V.}\ \bibnamefont {Meter}}, \bibinfo
  {author} {\bibfnamefont {A.}~\bibnamefont {Aspuru-Guzik}}, \ and\ \bibinfo
  {author} {\bibfnamefont {Y.}~\bibnamefont {Yamamoto}},\ }\href {\doibase
  10.1088/1367-2630/14/11/115023} {\bibfield  {journal} {\bibinfo  {journal}
  {New Journal of Physics}\ }\textbf {\bibinfo {volume} {14}},\ \bibinfo
  {pages} {115023} (\bibinfo {year} {2012})}\BibitemShut {NoStop}%
\bibitem [{\citenamefont {Fowler}()}]{Fowler2012}%
  \BibitemOpen
  \bibfield  {author} {\bibinfo {author} {\bibfnamefont {A.}~\bibnamefont
  {Fowler}},\ }\href@noop {} {\ }\Eprint {http://arxiv.org/abs/1210.4626}
  {arXiv:1210.4626} \BibitemShut {NoStop}%
\end{thebibliography}%

\end{document}